\documentclass[osajnl,a4paper,twocolumn]{revtex4}  
\usepackage{hyperref}

\usepackage{graphicx}

\begin{document}

\title{Noise and Signal scaling factors in Digital Holography  in weak illumination: relationship with Shot Noise.}

\author{M. Lesaffre$^1$, N. Verrier$^1$ and M. Gross$^2$}

\address{$^1$ Institut Langevin: UMR 7587 CNRS, ESPCI, ParisTech; 1 rue Jussieu, 75005 Paris, France\\
 $^2$ Laboratoire Charles Coulomb - UMR 5221 CNRS-
Universit\'{e} Montpellier 2. Place Eug\`{e}ne Bataillon, 34095 Montpellier, France}

\email{michel.gross@univ-montp2.fr}

\begin{abstract}
We have performed off axis heterodyne holography  with very weak illumination by   recording  holograms of the object with and without object  illumination  in the same acquisition run.  We have experimentally studied , how the  reconstructed image signal (with illumination) and  noise background (without)   scale with the holographic acquisition and reconstruction parameters that are the number of frames, and  the number of pixels of the reconstruction spatial filter. The first parameter is related to the frequency bandwidth of detection in time, the second one to the bandwidth in space.  The  signal to background ratio varies roughly like the inverse of the bandwidth in time and space. We have also compared the noise   background  with the theoretical shot noise background  calculated by Monte Carlo simulation. The experimental and Monte Carlo noise background agree very well together.
\end{abstract}

\maketitle

ocis{ 090.0090,0.090.0995, 120.2880}

\section{Introduction}

Demonstrated  by Gabor \cite{Gabor49} in the early 50's, the purpose of holography is to record, on a 2D detector, the phase and the amplitude of the light shining from an object under coherent illumination. Since a thin film does not provide a direct access to the recorded data, the holographic film has been  replaced by 2D electronic detector in digital holography \cite{macovski1971considerations}, whose main  advantage is to perform the data acquisition and the holographic reconstruction numerically \cite{goodman1967digital, schnars1994direct}.
Off-axis holography \cite{leith1965microscopy} is the oldest and the simplest
configuration adapted to digital holography \cite{Schnars_Juptner_94,schnars1994direct,kreis1998principles}. In off-axis digital holography, as well as in photographic plate holography, the reference or local oscillator (LO) beam is angularly tilted with respect to the object observation axis. It is then possible to record, with a single hologram, the two quadratures of the object's complex field. However, the
object field of view is reduced, since one must avoid the overlapping of the image with the conjugate image alias. Phase-shifting digital holography, which has been later introduced \cite{Yamaguchi1997}, records several images with a different phase for the LO beam. It is then possible to obtain the two quadratures of the field in an in-line configuration even though the conjugate image alias and the true image overlap, because aliases can be removed by taking image differences.

With the development of CCD camera techniques, digital holography became a fast-growing research field that has drawn increasing attention \cite{doval_2000,Schnars_2002}. Off-axis holography has been applied recently to particle \cite{pu2004intrinsic}, polarization \cite{colomb2002polarization}, phase contrast \cite{cuche1999digital}, synthetic aperture \cite{Massig_2002}, low-coherence \cite{ansari2001elimination,massatsch2005time} and microscopic \cite{massatsch2005time,Marquet_2005,atlan2008heterodyne}  imaging. Phase-shifting holography has been applied to 3D \cite{zhang1998three,nomura2007polarization}, color \cite{yamaguchi2002phase}, synthetic aperture \cite{Leclerc2001}, low-coherence \cite{tamano2006phase}, surface shape \cite{yamaguchi2006surface} and microscopic \cite{Zhang_1998,yamaguchi2001image} imaging.

We have developed an alternative phase-shifting digital holography technique that uses a frequency shift of the reference beam to continuously shift the phase of the recorded interference pattern \cite{Leclerc2000}.

More generally, our setup can be viewed as a multipixel heterodyne detector that is able of
recording the complex  field $E$ scattered by an object in all pixels of the CCD camera in parallel.  Because the holographic signal results in the interference of the object signal complex field $E$ with a reference (LO) complex field $E_{LO}$, whose amplitude can be  much larger (i.e. $|E_{LO}|\gg |E|$), the holographic detection  benefits of  ''heterodyne gain'' (i.e. $|E ~E^*_{LO}|\gg
|E|^2$), and is thus well suited to detect weak signal fields $E$.

By performing phase shifting  heterodyne holography in an off-axis configuration, the holographic signal can be frequency shifted both in time and space.  It is then  possible perform a very efficient filtering  of the holographic signal  in space and time. This double filtering combined with heterodyne gain open the way to holography of weak  object fields $E$ with ultimate  sensitivity  \cite{gross2007digital}. This type of holography has been used (i) to  detect and analyse the laser Doppler spectrum of the extremely  weak optical signal that is transmitted and scattered by a human breast in vivo ($<1/10$ photoelectron per pixel) \cite{gross2005heterodyne}, (ii)  to image, localize and analyse the motion of metallic nano particules \cite{warnasooriya2010imaging,verpillat2011dark}, and (iii)  to analyse vibration of very small amplitude \cite{gross2003shot,absil2010photothermal,psota2012comparison} by detecting the holographic signal on the vibration  sideband \cite{joud2009imaging}.

Since at low signal the prevalent and the limiting noise is shot noise \cite{davenport1958introduction}, it is important to study how shot noise limits the image quality in digital holography. Charri\`{e}re et al., have studied, with simulated and experimental data, how shot noise influences the quality of the phase images, in particular in the case of short exposure time \cite{charriere2006sni,charriere2007isn}. On the other hand, Verpillat et al.  have calculated, in the weak illuminated case, the shot noise limits for amplitude images \cite{verpillat2010digital}, and  have shown  that the shot noise background can be easily calculated by Monte Carlo simulations.

In this paper, we have considered amplitude images by performing off axis heterodyne holography  in very weak illumination  conditions. We have modified our experimental setup to record holograms with and without illumination of the object (but with reference beam) in the same acquisition run. We have measured the noise background  without illumination and compared it with the signal obtained with illumination. We have studied experimentally,   how the signal and the noise scales, on amplitude  reconstructed images,  with the two mains parameters of the holographic acquisition and reconstruction  that are the number of frames, and  the number of pixels of the reconstruction spatial filter \cite{cuche2000spatial}. The first parameter is related to the detection   frequency bandwidth in time, the second one to the bandwidth in space.  The expected scale factor values are retrieved: the signal to noise background ratio varies roughly like the inverse of the   bandwidth in time and space. Finally, we have compared the noise background observed in experiments, with the shot noise background  calculated by the Verpillat et al. Monte Carlo method. The experimental and Monte Carlo noise background agree very well together. This demonstrates that our weak illumination holography is shot noise limited.

\section{Experimental Setup and Numerical Reconstruction}

\begin{figure}[]
\begin{center}
\includegraphics[width = 6 cm,keepaspectratio=true]{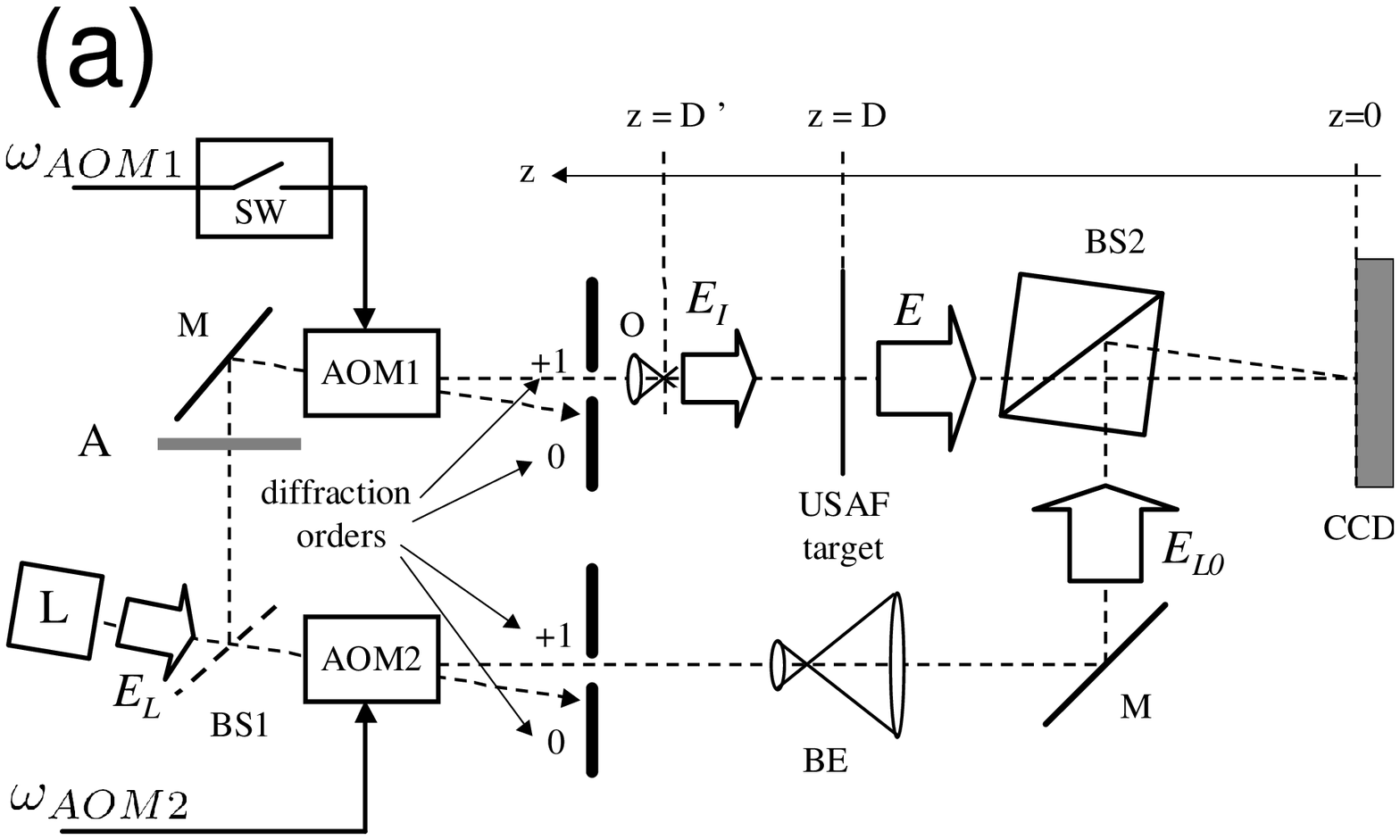}\\
\includegraphics[width = 6 cm,keepaspectratio=true]{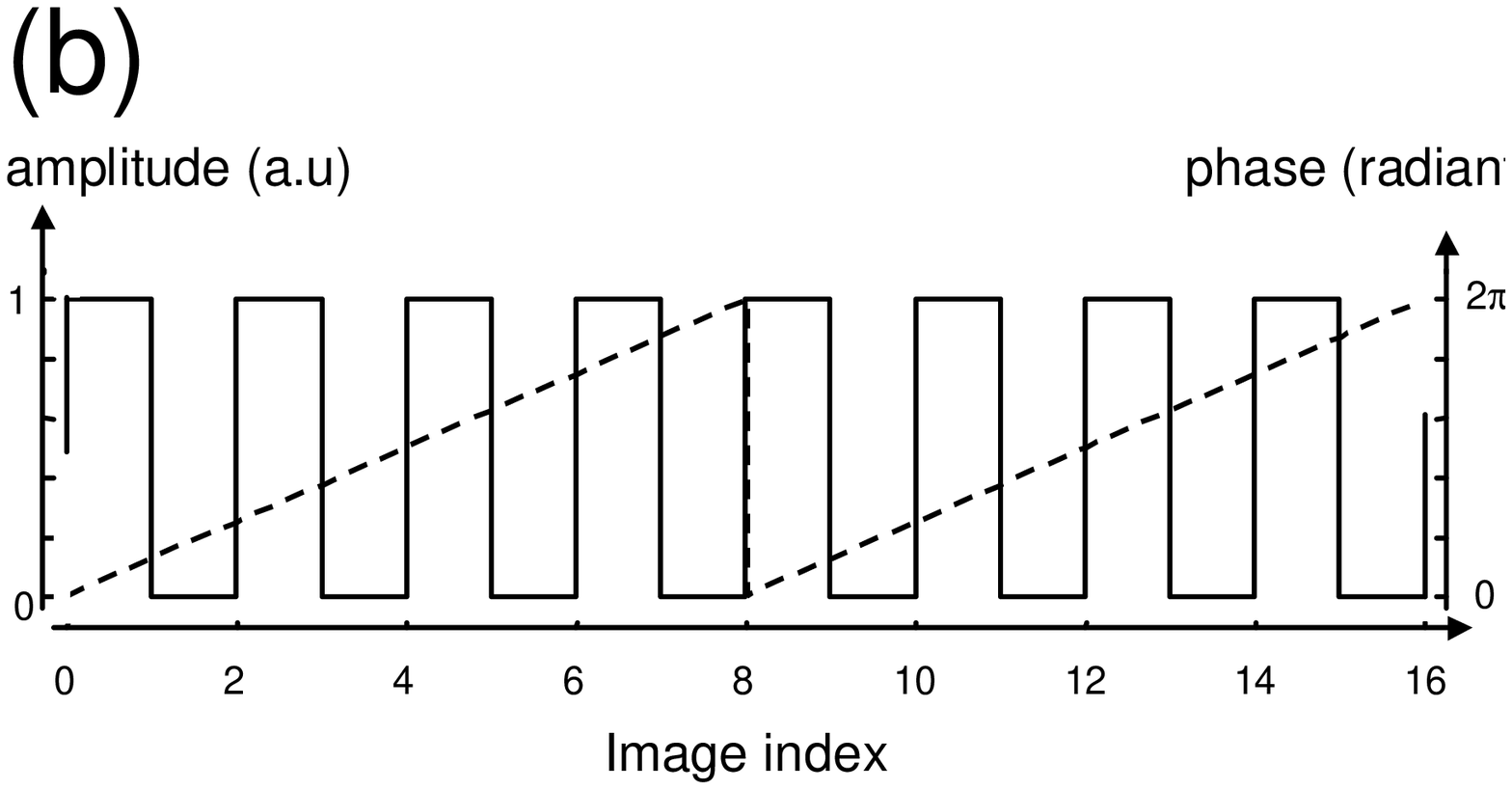}
\caption{(a) Digital holography setup. L: diode laser; BS1, BS2:
beam splitters; AOM1 and AOM2: acousto-optic modulators; O:
objective; BE: beam expander;  M: mirror; A: light attenuator. USAF:
transmission USAF target to image. CCD : CCD camera. (b)
Illumination beam amplitude (solid line) and phase (dashed line) as
a function of the frame index.} \label{fig_setup}
\end{center}
\end{figure}

The experimental setup is shown in Fig. \ref{fig_setup}(a). The light source is a $\lambda = 532$ nm, 50 mW laser (SLIM-532: Oxxius SA). The laser beam (complex field $E_L$, frequency $\omega_L$) is split into an illumination beam ($E_I$, $\omega_I$) and a reference local oscillator (LO) beam ($E_{LO}$, $\omega_{LO}$). The LO beam is enlarged by a beam expander BE, and is tilted by an angle $\theta \sim 1^\circ$ with respect to the observation axis of the CCD camera.

The illumination beam is enlarged by a short focal length (5 mm) objective O, located at a distance $D'-D\simeq 63$ cm in front of an United States Air Force (USAF) target. The light transmitted by this object is superimposed on the LO beam. The interference pattern is recorded by the CCD camera (PCO Inc. Pixelfly digital camera: 12 bit, frame rate $\omega_{CCD}$=12.5 Hz, exposure time T = 50 ms, with 1280 $\times$ 1024 pixels of 6.7 $\times $ 6.7 $\mu$m) located at $D=32.4$ cm downstream. The signal of the camera  measured in  Digital Count (DC)
units  (0 to 4095) is  multiplied by the camera gain ($G=4.41$ e/DC as calibrated by PCO Inc.) to be converted in photoelectron (e) units. A set of optical attenuators  A (gray neutral filters) is used to reduce the intensity of the illumination beam.

Two acousto-optic modulators (Crystal Technology: $\omega_{AOM1,2}\simeq 80$ MHz) are used to adjust $\omega_I$ and $\omega_{LO}$:
\begin{eqnarray}
  \omega_I&=&\omega_L+\omega_{AOM1}\\
\nonumber  \omega_{LO}&=&\omega_L+\omega_{AOM2}
\end{eqnarray}
such way the phase of the $E E_{LO}^*$ term is shifted by $\pi/4$ from one CCD frame to the next (8 phases detection). We have thus:
\begin{eqnarray}
\omega_{LO}-\omega_I&=&\omega_{AOM2}-\omega_{AOM1} = \omega_{CCD}/8
\end{eqnarray}
In addition, the $\omega_{AOM1}\simeq 80$ MHz signal that drives  AOM1 is
modulated in amplitude at $\omega_{CCD}/2$ through the switch SW, so that the illumination beam is turned on and off from one CCD frame to the next. The amplitude and phase (relative to the LO beam) of the illumination beam are displayed in Fig.\ref{fig_setup} (b). Even frames are recorded with the illumination beam on the object, whereas odd frames are recorded without.  We have then recorded sequences of $64$ successive CCD frames ($I_0$ to $I_{63}$) that are analyzed to reconstruct the intensity images of the USAF target.

Compared with the  heterodyne holography setup  of ref \cite{gross2007digital}, the Fig.\ref{fig_setup} setup exhibits two main differences. First, the AOMs acts on both the illumination  and  local oscillator arms (and not  on the local oscillator arm only). It is thus possible to put the illumination attenuator A upstream with respect to AOM1. The  parasitic light (of frequency $\omega_L$)  that is diffused by the optics upstream with respect to the attenuator A is then frequency shifted by $80 $ MHz with respect to the detected  frequency ( $\simeq \omega_{LO}$). This parasitic light is thus not detected, and the illumination leaks are  reduced.

Second, the electronic switch turns on and off the illumination in such a way that the even frames are recorded with illumination of the sample, and the odd frames without (but the LO beam remains for both even and odd frames). In the following "with illumination" means thus with illumination and LO beams on, while "without illumination" means with illumination beam off and LO beam on.   The frequency shift of the LO is here $\omega_{CCD}/8$ (and not $\omega_{CCD}/4$) to record standard 4 phases hologram with and without object  illumination: the phase shift between successive even  (or odd) frames  is therefore $\pi/2$.

We must notice that the combination of the electronic switch with the 8 phases detection is not needed for making 4 phase phase shifting holography. It has been used here to be able  to extract from the same sequence of data different types of holograms. With the even frames (or odd frames),  one can perform 4 phase holography with or without illumination of the sample. This can  be done with 4, 8, 16 or 32 frames.  One can also consider only the 0$^\textrm{th}$ and 4$^\textrm{th}$ (or 1$^\textrm{th}$ and 5$^\textrm{th}$) frames to make 2 phase holography with or without object  illumination.  By using the 0$^\textrm{th}$ frame and making the difference with the 1$^\textrm{st}$ frame  one can  also perform single shot off-axis holography with  suppression of the DC term (as proposed by Kreis et al. \cite{kreis1997suppression}). In the following we have analyzed the reconstructed images corresponding to these different types of hologram  to better understand how the signal and the noise depend on the  number of frames.

Many numerical methods can be used to reconstruct the image of  the object  \cite{picart2008general,verrier2011off}.
The most common methods  are the  Schnars et al. methods \cite{Schnars_Juptner_94} that involves a single discrete Fast Fourier Transform (FFT), and the angular spectrum method that involves two FFT \cite{Leclerc2000,yu2005wavelength}.
In the last case, it is possible to use the Cuche et al. method \cite{cuche2000spatial} to  perform a spatial filtering of the pertinent holographic signal.

We have consider here the angular spectrum method with spatial filtering, and we have  studied how the holographic signal and  noise scale with the number of frames, and  with the area of the   spatial  filter.
\begin{figure}[]
\begin{center}
\includegraphics[width =7.5 cm,keepaspectratio=true]{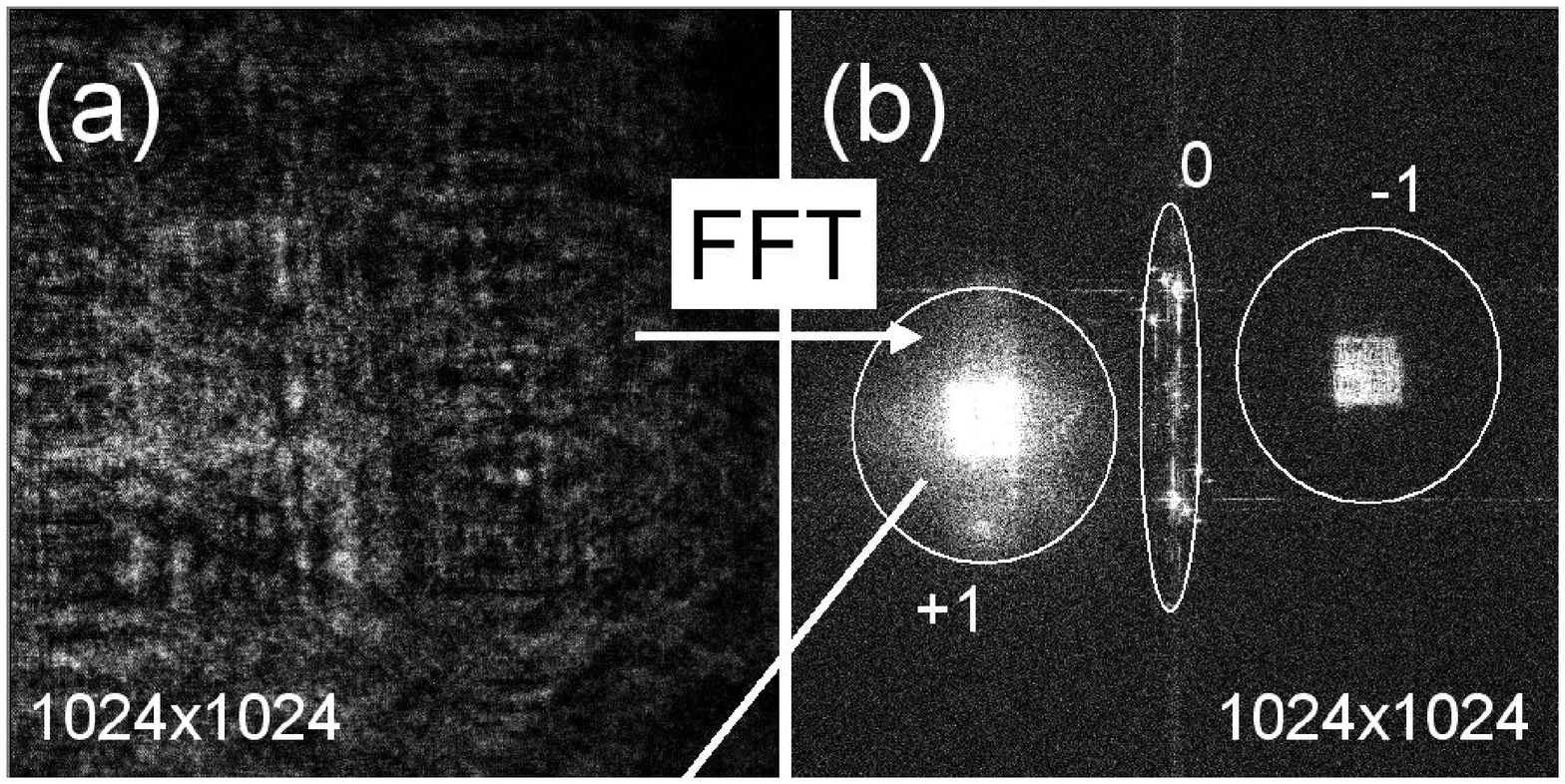}\\
\includegraphics[width =7.5 cm,keepaspectratio=true]{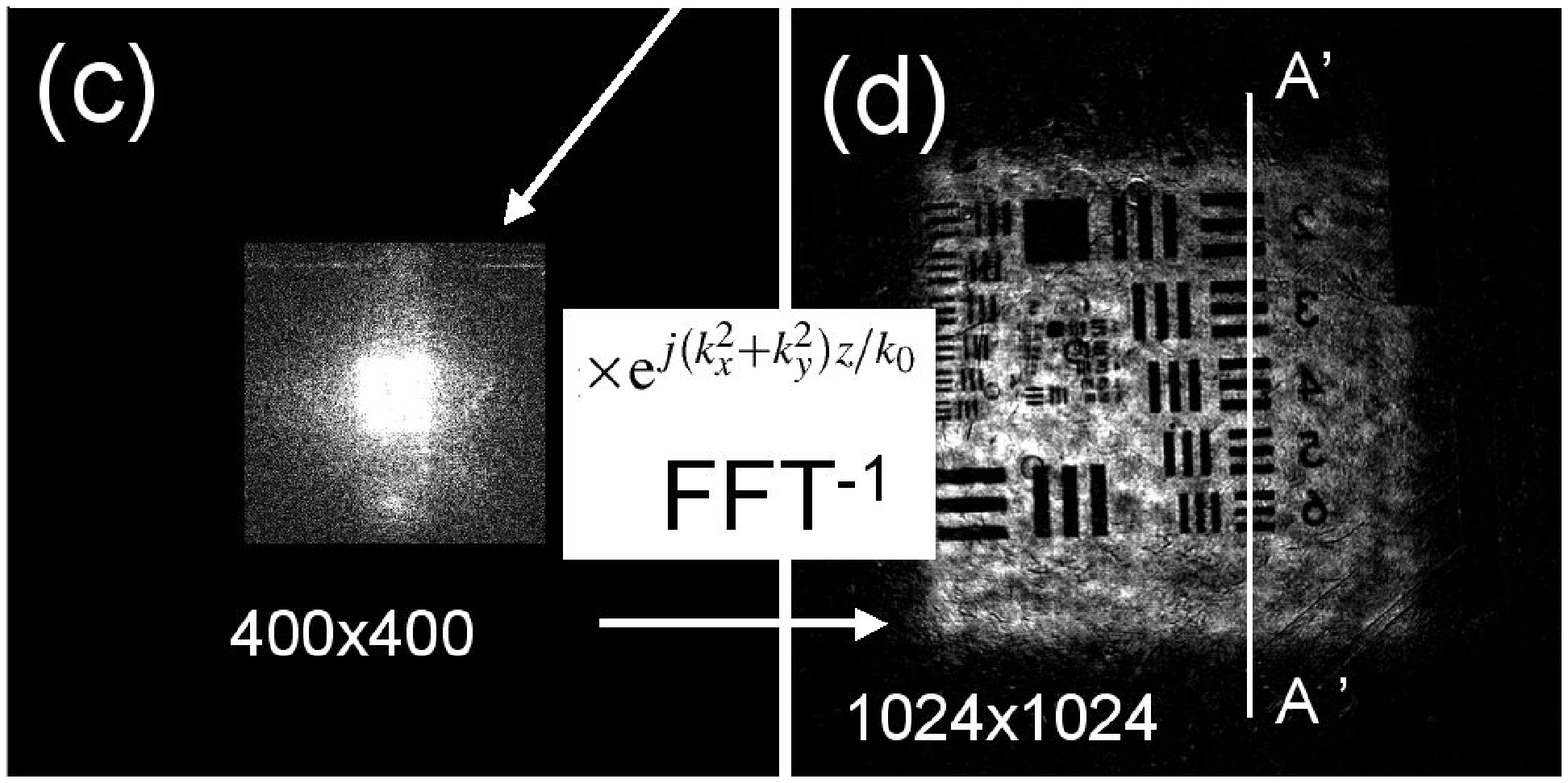}\\
\caption{ (a) Intensity signal   $|H(x,y,z=0)|^2$ detected by the CCD camera. (b,c) k-space intensity signal
$|\tilde H(k_x,k_y,z=0)|^2$ without (b) and  k-space filtering (c) with two  filter sizes
($400\times 400$. (d) Reconstructed images $|H(x,y,z=D)|^2$ of the
USAF target. Display is in arbitrary log scale. }
\label{fig2_Fig2_recons}
\end{center}
\end{figure}

To illustrate the reconstruction (see Fig. \ref{fig2_Fig2_recons}), we have considered a phase shifting complex hologram $H$ with 4 phases: $H=(I_0-I_2)+j(I_1-I3)$, where $I_0$, $I_1$  ....$I_3$ are the CCD signals recorded on 4 successive frames and $j^2=-1$.
The intensity of the CCD plane  complex hologram $|H(x,y,0)|^2$ (where $x$ and $y$ are the coordinate of the CCD pixels) is displayed on Fig.\ref{fig2_Fig2_recons} (a). The reciprocal space hologram $\tilde{H}(k_x,k_y,0)$ (where $k_x$ and $k_y$ are the coordinates of the reciprocal space)  is calculated from the CCD plane hologram $H(x,y,z=0)$  by a Fast Fourier Transform (FFT):
\begin{equation}\label{eq_FFT1}
\tilde{H}(k_x,k_y,0) = {\rm FFT} \left[H(x,y,0) \right]
\end{equation}
The reciprocal space hologram $\tilde{H}(k_x,k_y,0)$  is displayed on Fig. \ref{fig2_Fig2_recons} (b). The USAF target information lies in the bright zone in the left hand side of Fig. \ref{fig2_Fig2_recons} (b)  that corresponds to order +1 image. Because of the 4 phase detection, the  zero order and order -1 images are lower in intensity. There are nevertheless still visible because of logarithmic scale display.
To  select the   order +1 signal, we have here performed  spatial filtering with a $400\times 400$ pixels filter. A   $400\times 400$ pixels  square region centered on  order +1 is thus  cropped, and translated  to the centre of a zero-filled $1024 \times 1024$ calculation grid (zero padding) as shown on Fig.\ref{fig2_Fig2_recons}(c). The cropping operation performs a spatial filtering in the  Fourier space \cite{cuche2000spatial}, while the translation compensates for the off-axis geometry.

In the reciprocal space, the hologram $\tilde H$  is then multiplied by the quadratic phase factor that describes  the optical field propagation from the CCD plane $z=0$ to the USAF target plane $z=D$
\begin{equation}\label{eq_mult_phase_matrix}
\tilde{H}(k_x,k_y,z)=\tilde{H}(k_x,k_y,0) ~  \textrm{e}^{j (k_x^2+k_y^2) z/k_0},
\end{equation}
where $k_0=\sqrt{k_x^2+ k_y^2+ k_z^2}=2\pi/\lambda$.
The reconstructed image $H(x,y,D)$  is then obtained  by inverse FFT:
\begin{equation}\label{eq_FFT-1}
H(x,y,z)= {\rm FFT}^{-1} \left[\tilde{H}(k_x,k_y,z) \right]
\end{equation}
$H(x,y,D)$  is displayed on Fig.\ref{fig2_Fig2_recons}(d). Because the reconstruction involves 2  FFT, the pixel size is the same than in the CCD hologram  i.e.  6.7 $\times $ 6.7 $\mu$m.

\section{Experimental results }

We have recorded  series of 64 frames for different illumination levels of the USAF target, and we have reconstructed holographic images of the target by varying (i) the parity of the frames, (ii) the number of frames,  and (iii) the area of the spatial filter.

To analyze the results, we have quantified the noise in  different ways. First, intensity images have been   displayed in order to get  a qualitative idea of the noise. To make a more qualitative analysis, we have  performed a cut along a line crossing the black bars of the USAF target (like line AA' of Fig.\ref{fig2_Fig2_recons} (c)), and we have plotted the intensity along this cut. We   have averaged the pixel intensity over the whole reconstructed image, and we have plotted this average intensity as a function of the reconstruction parameter (parity, number of frames, and area of the spatial filter). We have finally compared the noise background with the one resulting from shot noise.

\subsection{Images and cuts}


We have first studied how the signal, and the noise depends on the number of frames recorded with and without signal. From the sequence of CCD frames $I_n$ (with $n=0$ to 63), we have built $n$ frame holograms $H_n$ with illumination (even frames) defined by:
\begin{eqnarray}\label{Eq_holograms_H}
 H_2 &=& I_0-I_4 \\
\nonumber  H_{4} &=&  (I_0-I_4)+j(I_2-I_6) \\
\nonumber  H_{4K} &=& \sum_{k=0}^{k=K-1}~(I_{8k}-I_{8k+4})+j(I_{8k+2}-I_{8k+6})
\end{eqnarray}
with $K=2,4,8$. In this equation $H_2$  a 2 phase phase-shifted hologram and $H_4$ a 4 phase holograms. $H_8$, $H_{16}$ and $H_{32}$ are 4 phase holograms recorded with 8, 16, and 32 frames.

Similarly to the  holograms $H_n$ constructed with $n$  even frames,  we have constructed the  holograms $H'_n$  with $n$ odd frames (i.e. without object  illumination) by replacing  in Eq. (\ref{Eq_holograms_H}) the even frames $I_{2n}$ by the  odd frames $I_{2n+1}$.
%
We have also consider the single illuminated frame hologram $H_{1,1}$ obtained by subtracting the reference frame $I_1$ recorded without   illumination of the object to the  frame $I_0$ recorded with illumination (the two indexes 1 and 1 correspond here to the number of even and odd  frames)~\cite{kreis1997suppression}:
\begin{eqnarray}\label{Eq_holograms_H11}
  H_{1,1} &=& I_0 - I_1
\end{eqnarray}

\begin{figure}[]
\begin{center}
\includegraphics[width =4.0 cm,keepaspectratio=true]{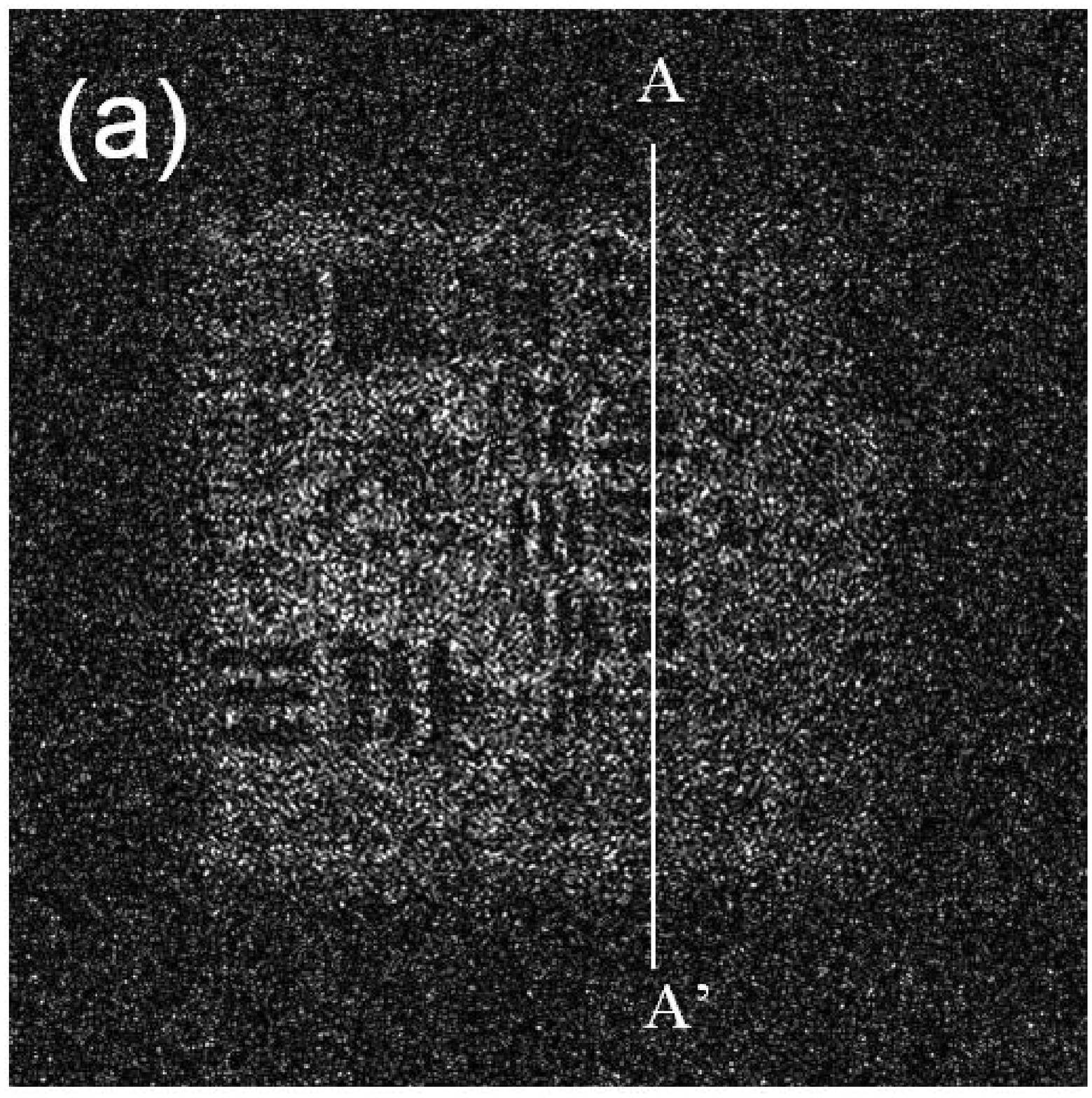}
\includegraphics[width =4.0 cm,keepaspectratio=true]{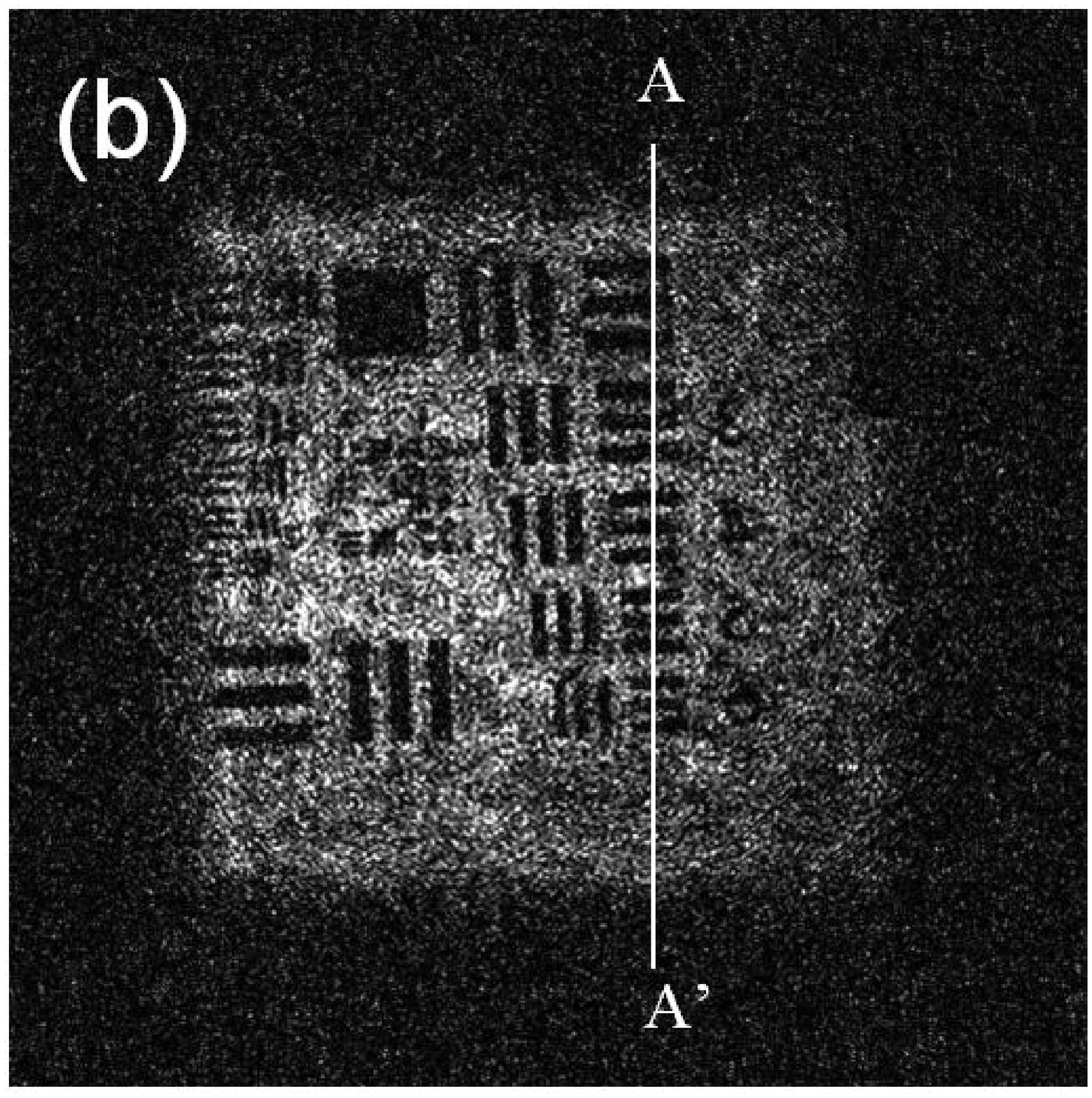}\\
\includegraphics[width =4.0 cm,keepaspectratio=true]{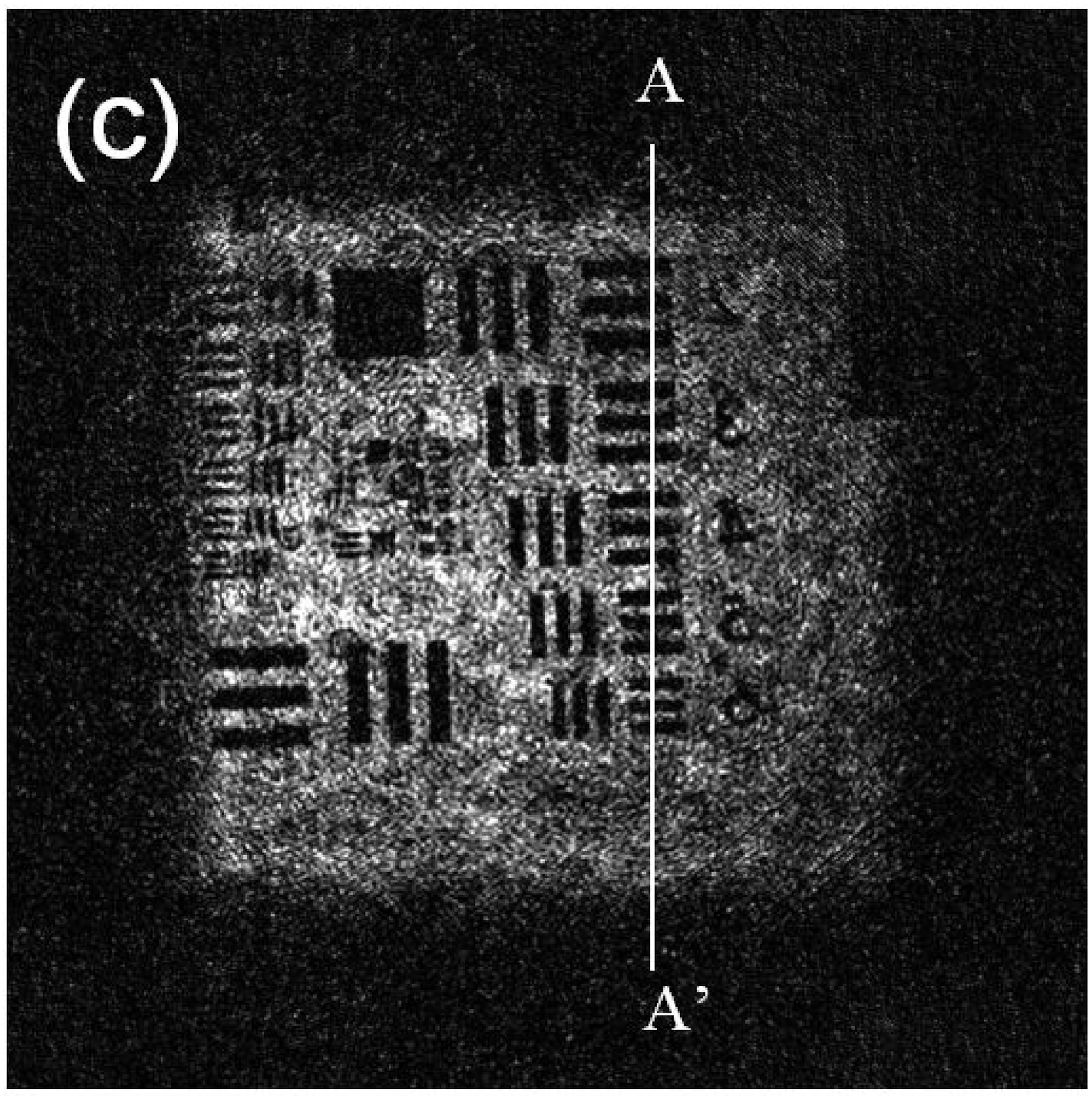}
\includegraphics[width =4.0 cm,keepaspectratio=true]{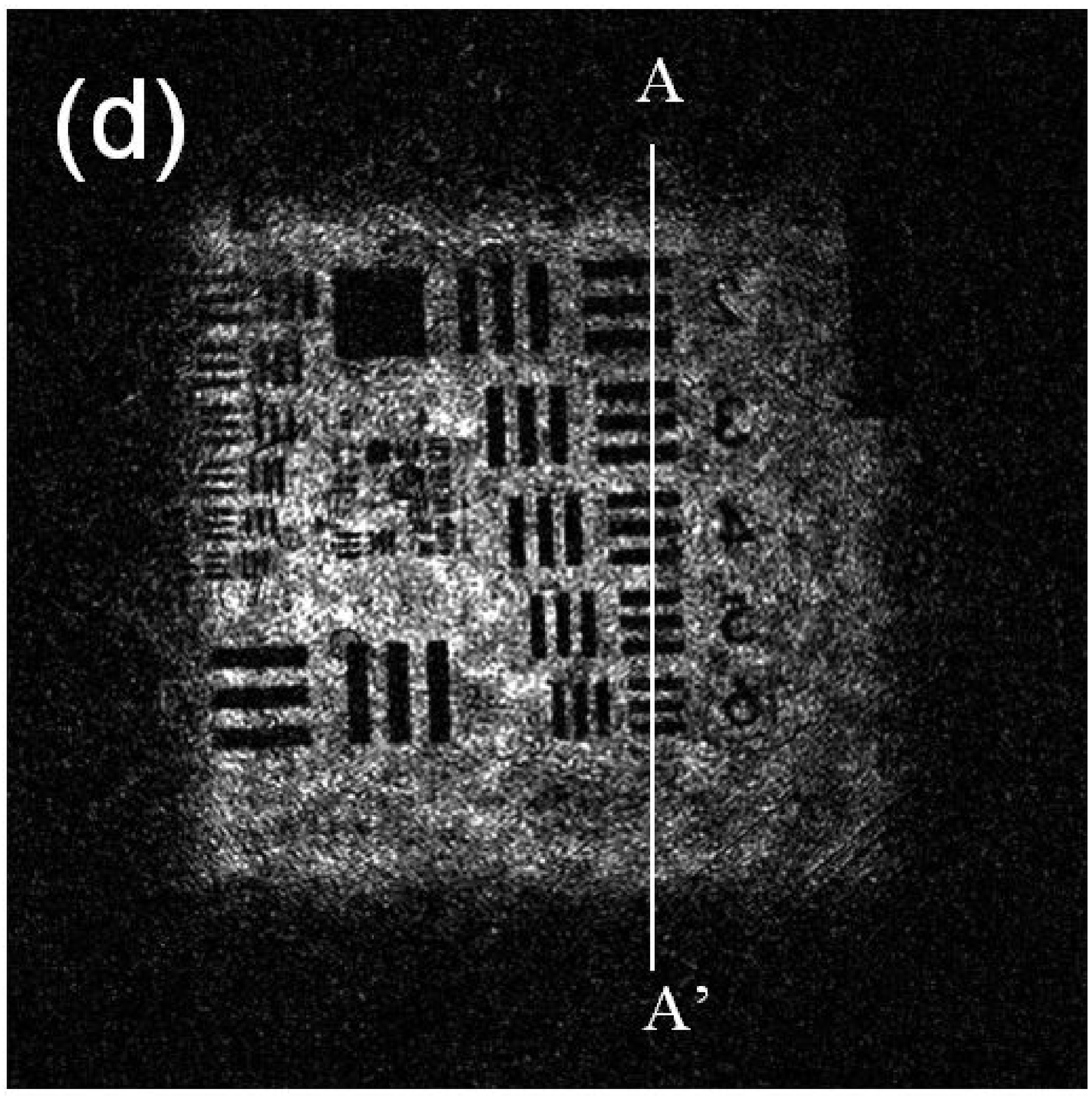}\\
\caption{Intensity image (i.e. $|H(x,y,D)|^2$) of the USAF target reconstructed from  $  H_{1,1}$ (a), $ H_{2}$ (b), $H_{4}$ (c), $ H_{8}$ (d).  Reconstruction with spatial filter $282\times 282$ pixels. }
\label{fig_images_2FFT_282_5}
\end{center}
\end{figure}

Figure \ref{fig_images_2FFT_282_5}  shows  reconstructed images of the USAF target illuminated at very low level (estimated to be 0.16 photon electron per pixel and per frame) with a spatial filter of $282\times 282$ pixels. Figure \ref{fig_images_2FFT_282_5} (a) is obtained from $  H_{1,1}$ (i.e. with one frame with and one frame without object  illumination). The Signal to Noise Ratio (SNR) of the holographic image is low, and the black bars of the target are barely visible (here, and in the remaining of this article, the term SNR must be understood, in a very qualitative way: SNR qualify the visual quality of either an image or a curve).  Figure \ref{fig_images_2FFT_282_5} (b,c) and (d)  show  the reconstructed images obtained from $H_2$,  $H_4$ and $H_8$ i.e. with 2, 4 and 8 illuminated frames. As expected the  SNR is becoming greater and greater.
\begin{figure}[]
\begin{center}
\includegraphics[width =4.0 cm,keepaspectratio=true]{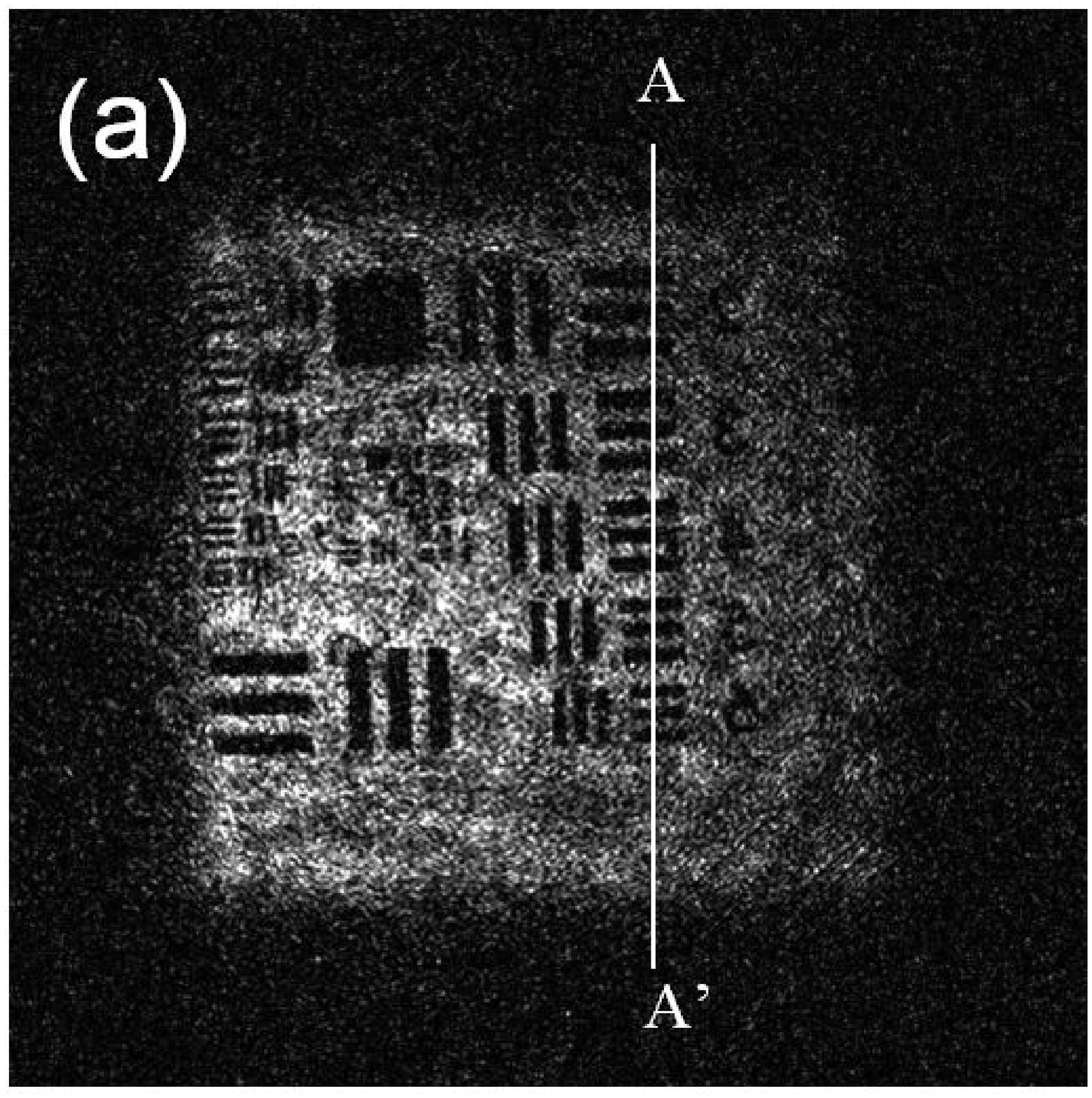}
\includegraphics[width =4.0 cm,keepaspectratio=true]{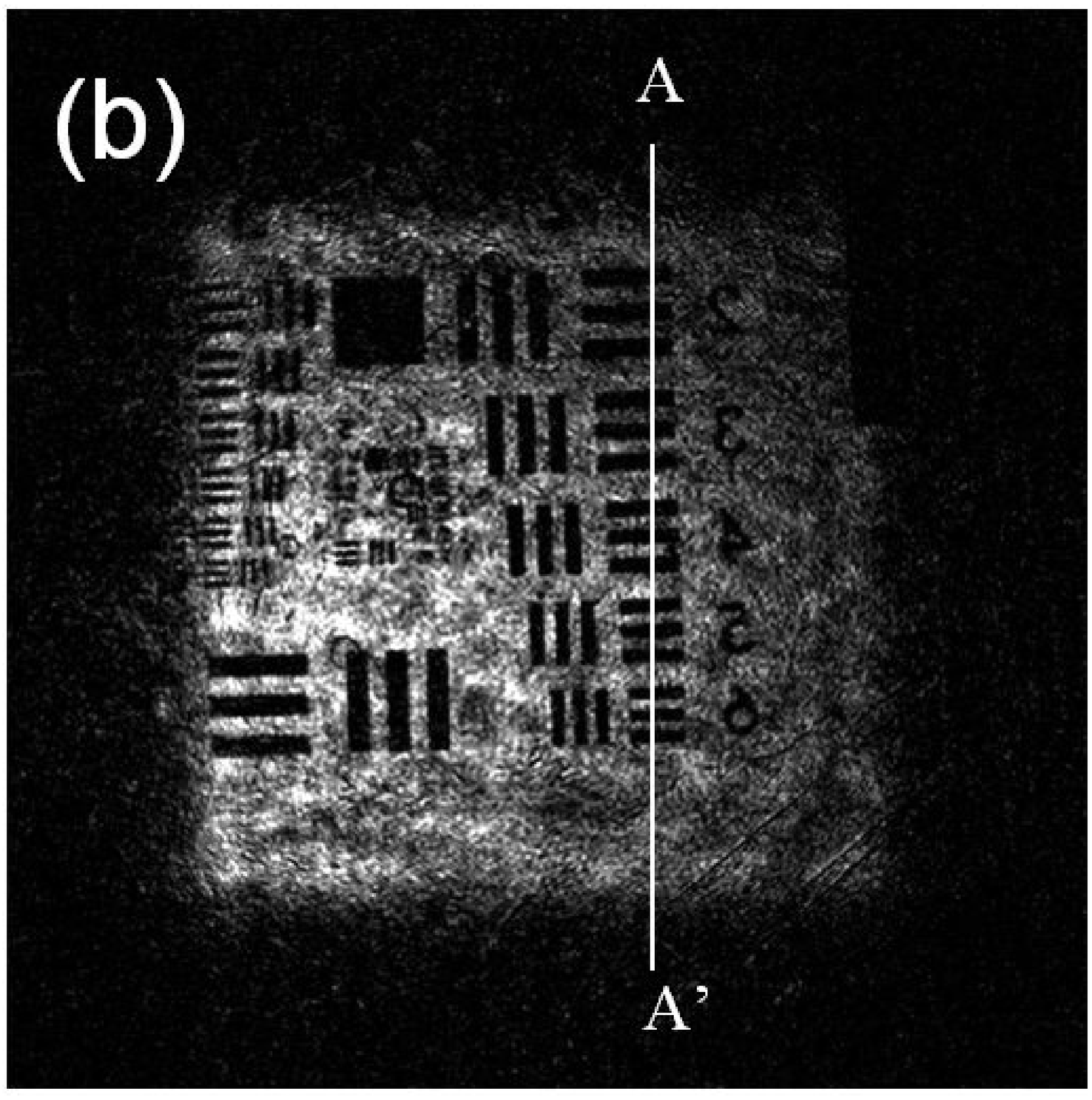}\\
\includegraphics[width =4.0 cm,keepaspectratio=true]{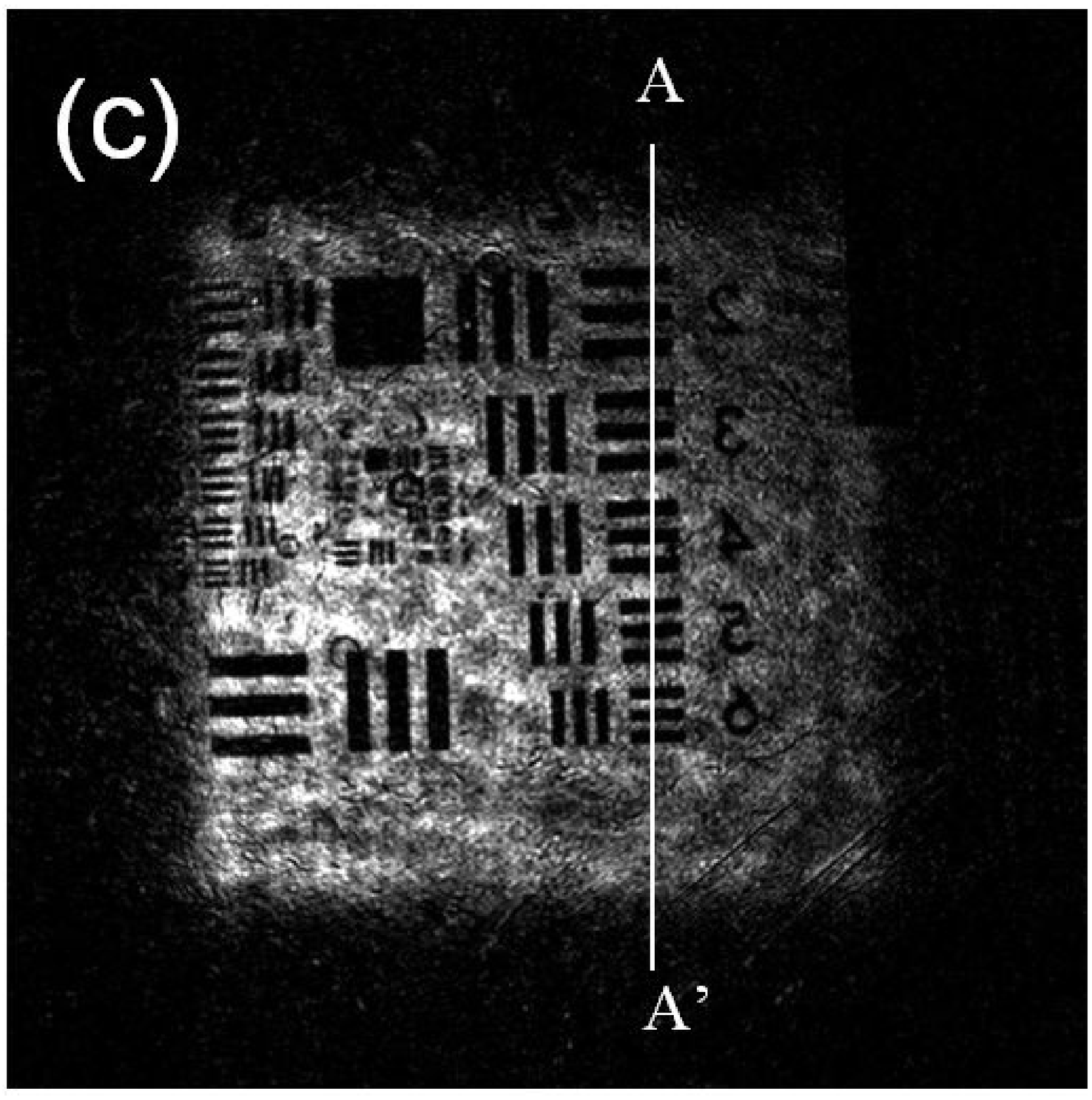}
\includegraphics[width =4.0 cm,keepaspectratio=true]{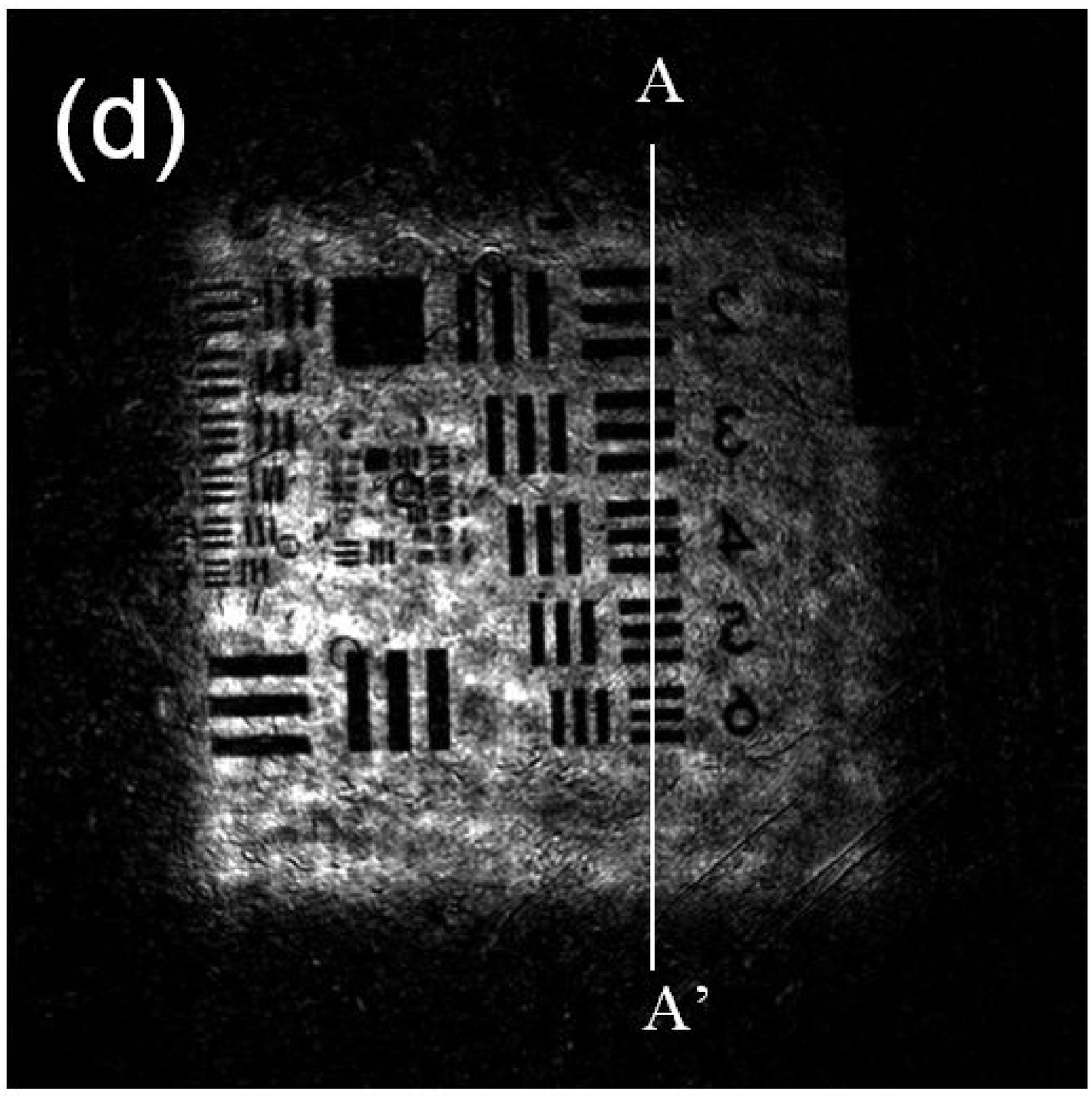}\\
\caption{Same as Fig.\ref{fig_images_2FFT_282_5} with higher illumination: intensity image (i.e. $|H(x,y,D)|^2$) of the USAF target reconstructed from  $  H_{1,1}$ (a), $ H_{2}$ (b), $H_{4}$ (c), $ H_{8}$ (d).  Reconstruction with spatial filter $282\times 282$ pixels. Vertical axis units are e$^2$ per pixel.}
\label{fig_images_2FFT_282_4}
\end{center}
\end{figure}
Figure \ref{fig_images_2FFT_282_4}  is similar to Fig.\ref{fig_images_2FFT_282_5}. It  shows the  reconstructed images at a slightly higher illumination level (i.e. 1.8 photon electrons per pixel and per frame). As expected, SNR is higher. The bars are visible on  all images i.e. on Fig. \ref{fig_images_2FFT_282_4} (a) to (d). As expected the  SNR is becoming greater and greater from Fig.\ref{fig_images_2FFT_282_4} (a) to (d) i.e. from 1  to 8  frames with illumination or  from $H_{1,1}$ to $H_8$.

\begin{figure}[]
\begin{center}
\includegraphics[width =4.2 cm,keepaspectratio=true]{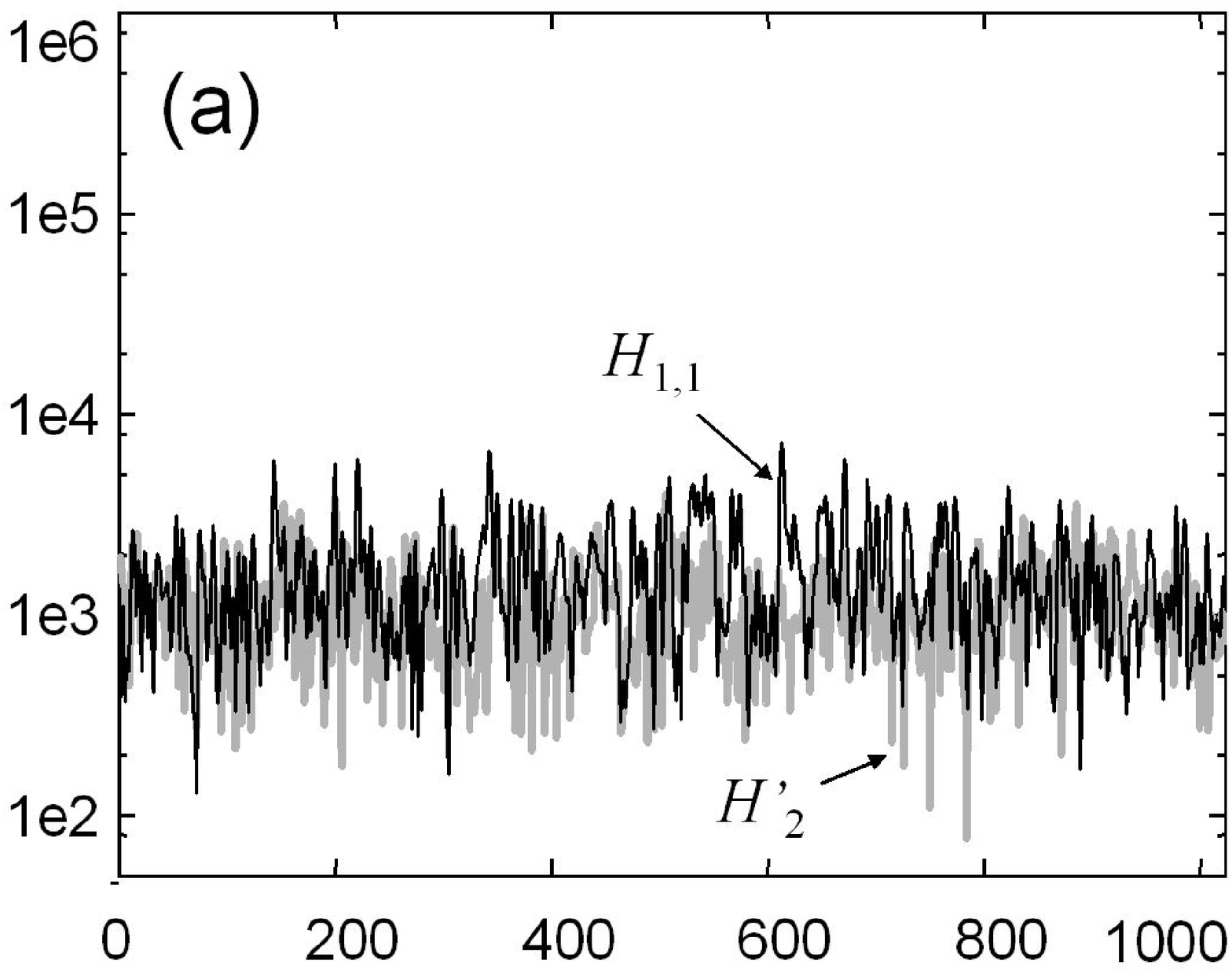}
\includegraphics[width =4.2 cm,keepaspectratio=true]{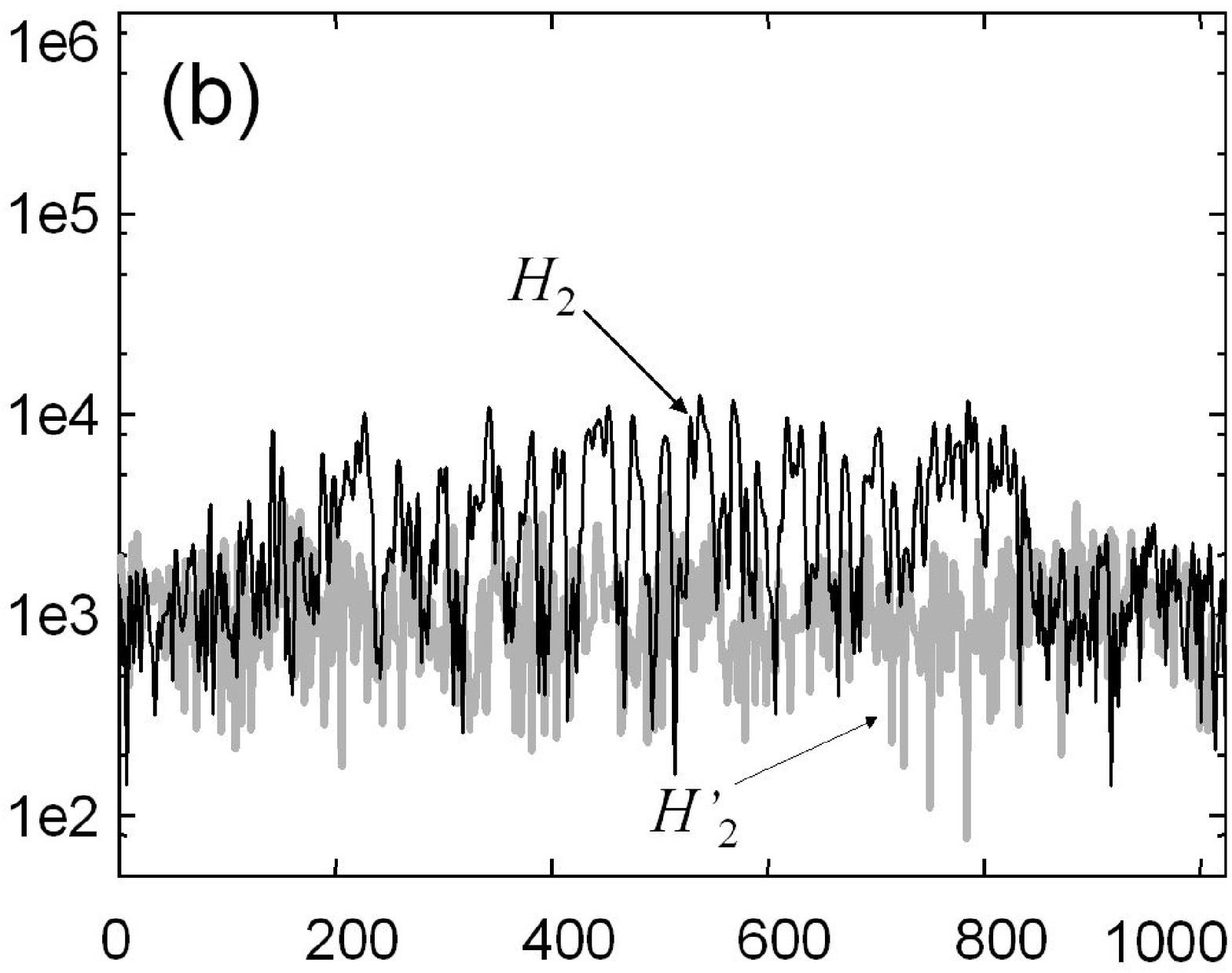}\\
\includegraphics[width =4.2 cm,keepaspectratio=true]{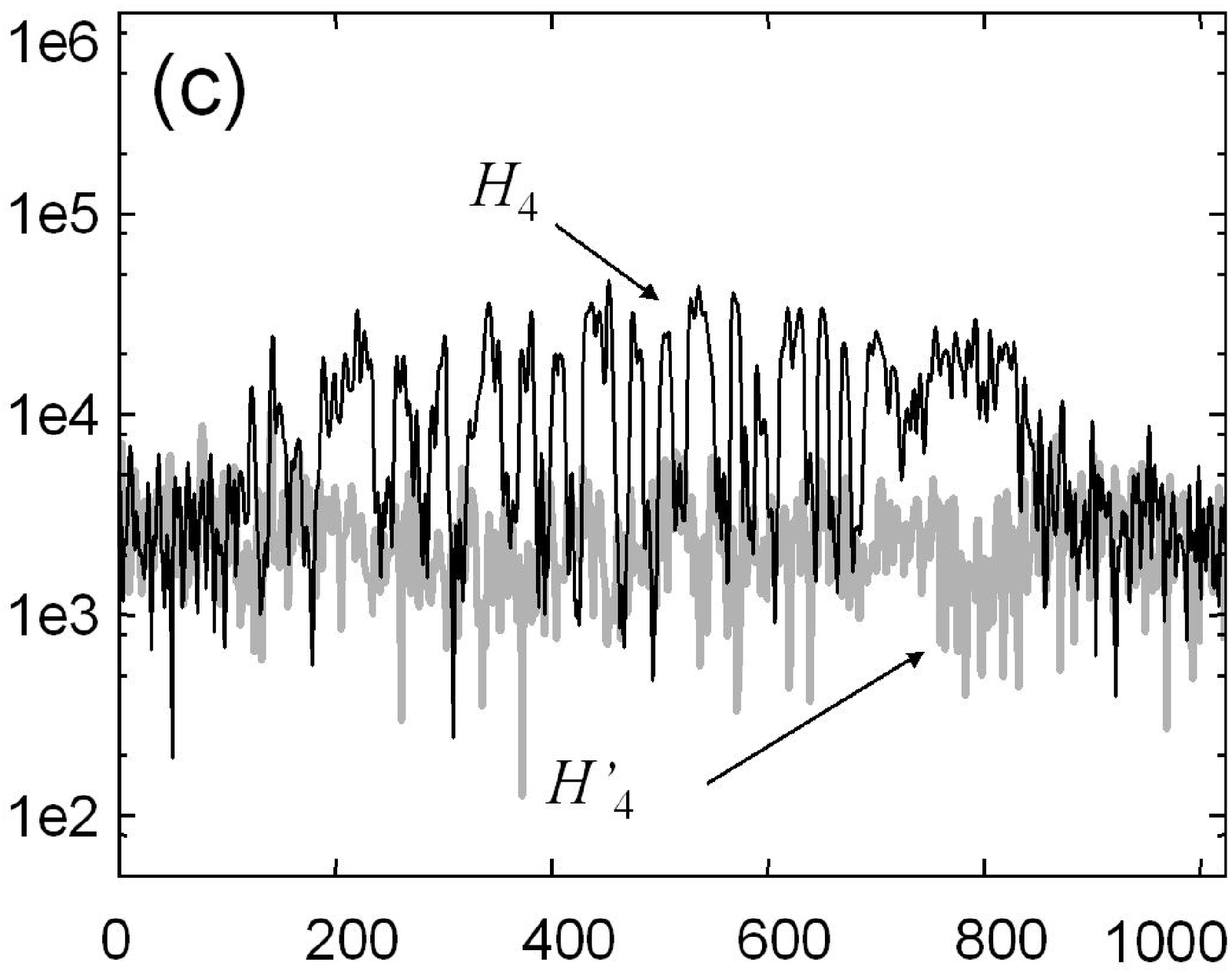}
\includegraphics[width =4.2 cm,keepaspectratio=true]{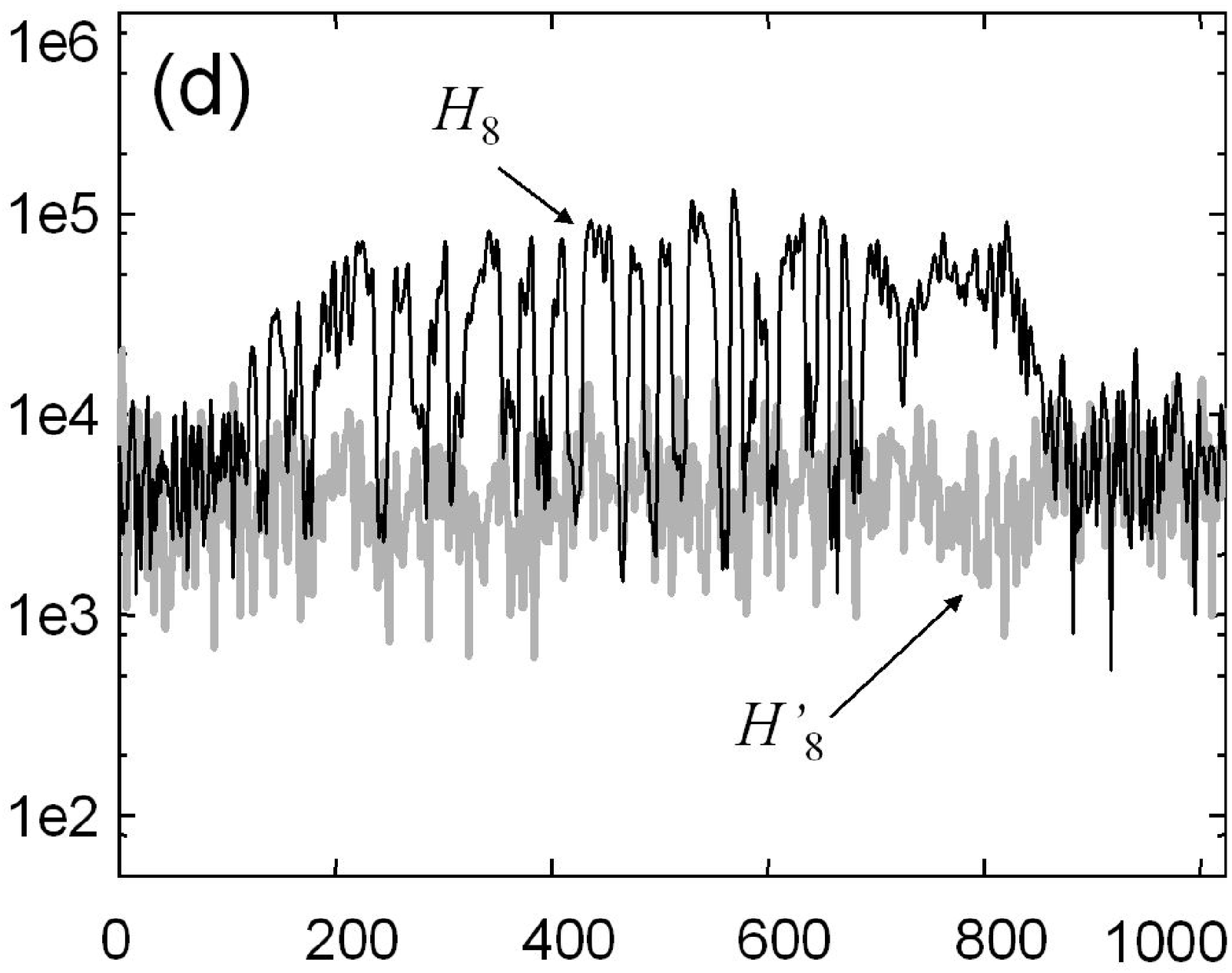}\\
\caption{Intensity (i.e. $|H(x,y,D)|^2$) along the  AA' white line of Fig.\ref{fig_images_2FFT_282_5}. The black curves correspond  $  H_{1,1}$ (a), $ H_{2}$ (b), $H_{4}$ (c) and $ H_{8}$ (d). The light grey curves in background  correspond  to  $  H'_{2}$ (a), $ H'_{2}$ (b), $H'_{4}$ (c) and $ H'_{8}$ (d). Reconstruction is made with a spatial filter of $282\times 282$ pixels. Vertical axis units are e$^2$ per pixel.}
\label{fig_cut_2FFT_282_5}
\end{center}
\end{figure}
\begin{figure}[]
\begin{center}
\includegraphics[width =4.2 cm,keepaspectratio=true]{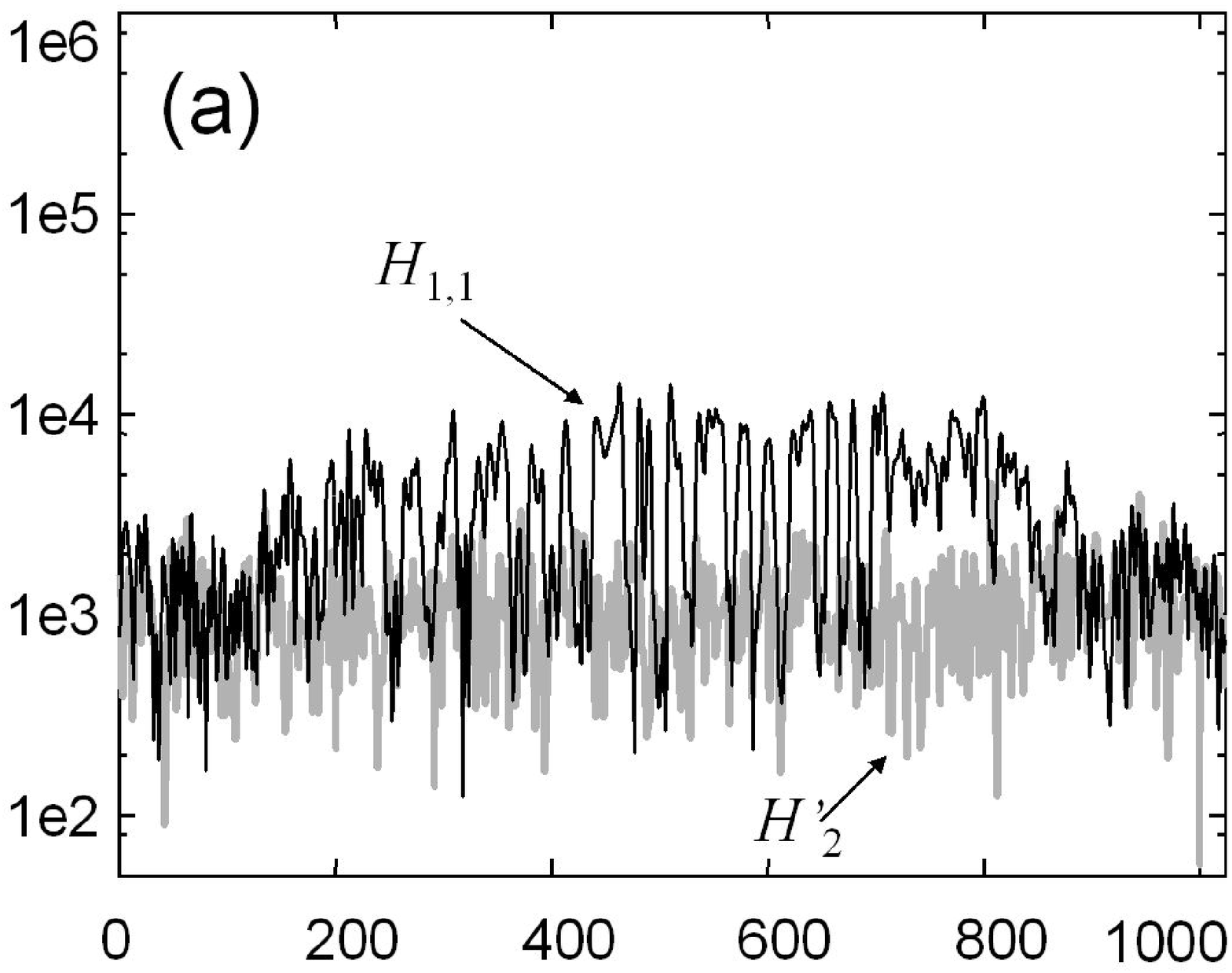}
\includegraphics[width =4.2 cm,keepaspectratio=true]{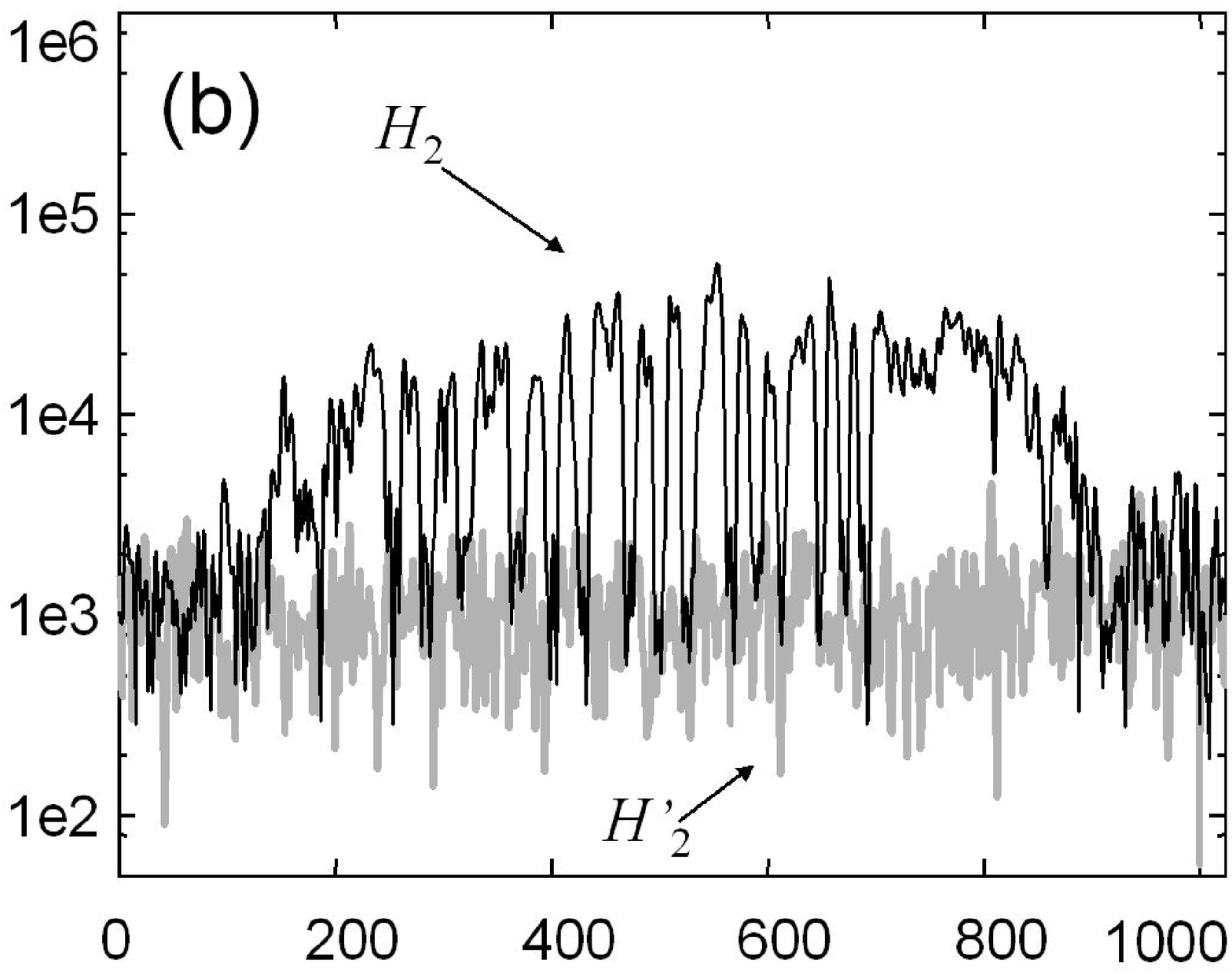}\\
\includegraphics[width =4.2 cm,keepaspectratio=true]{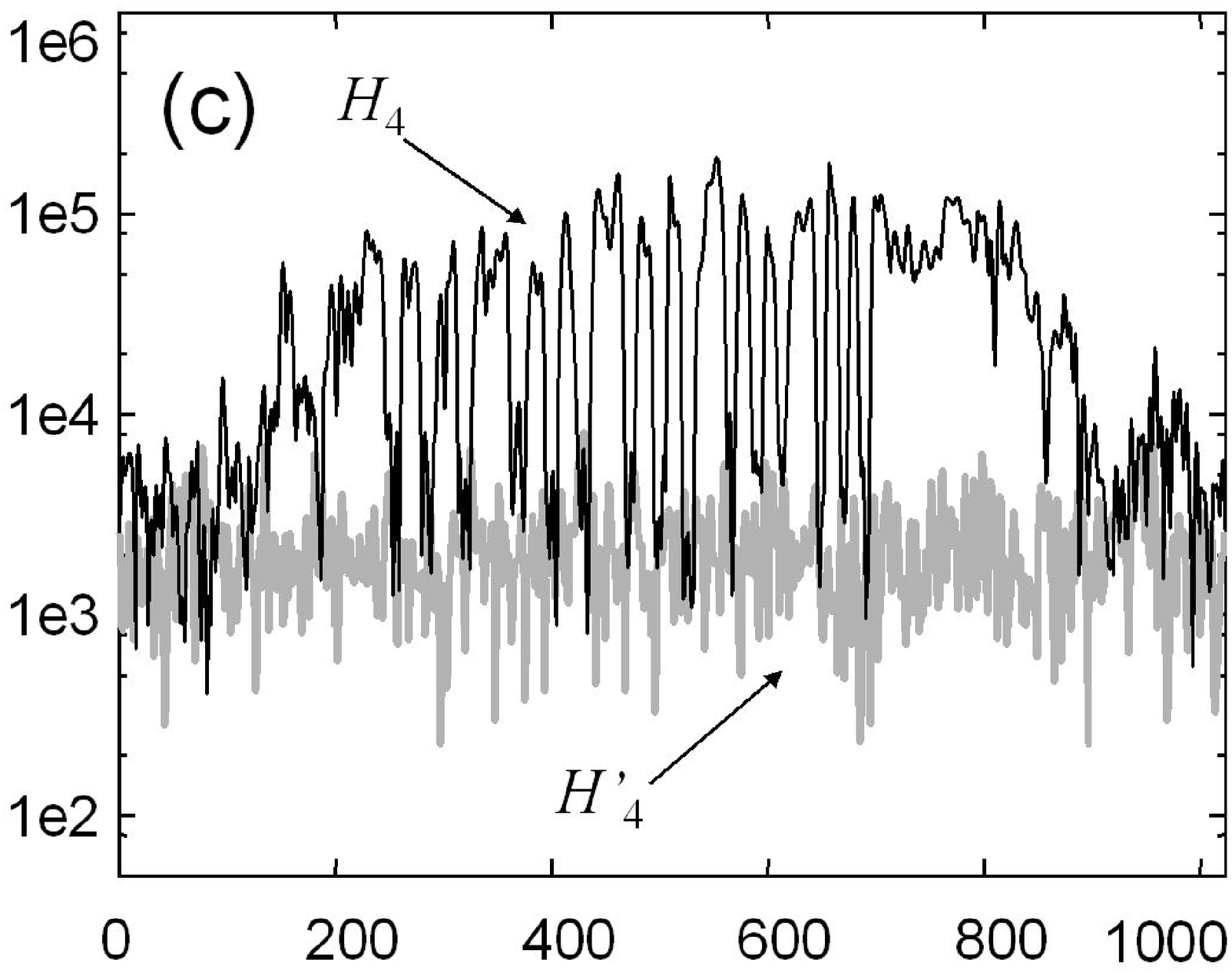}
\includegraphics[width =4.2 cm,keepaspectratio=true]{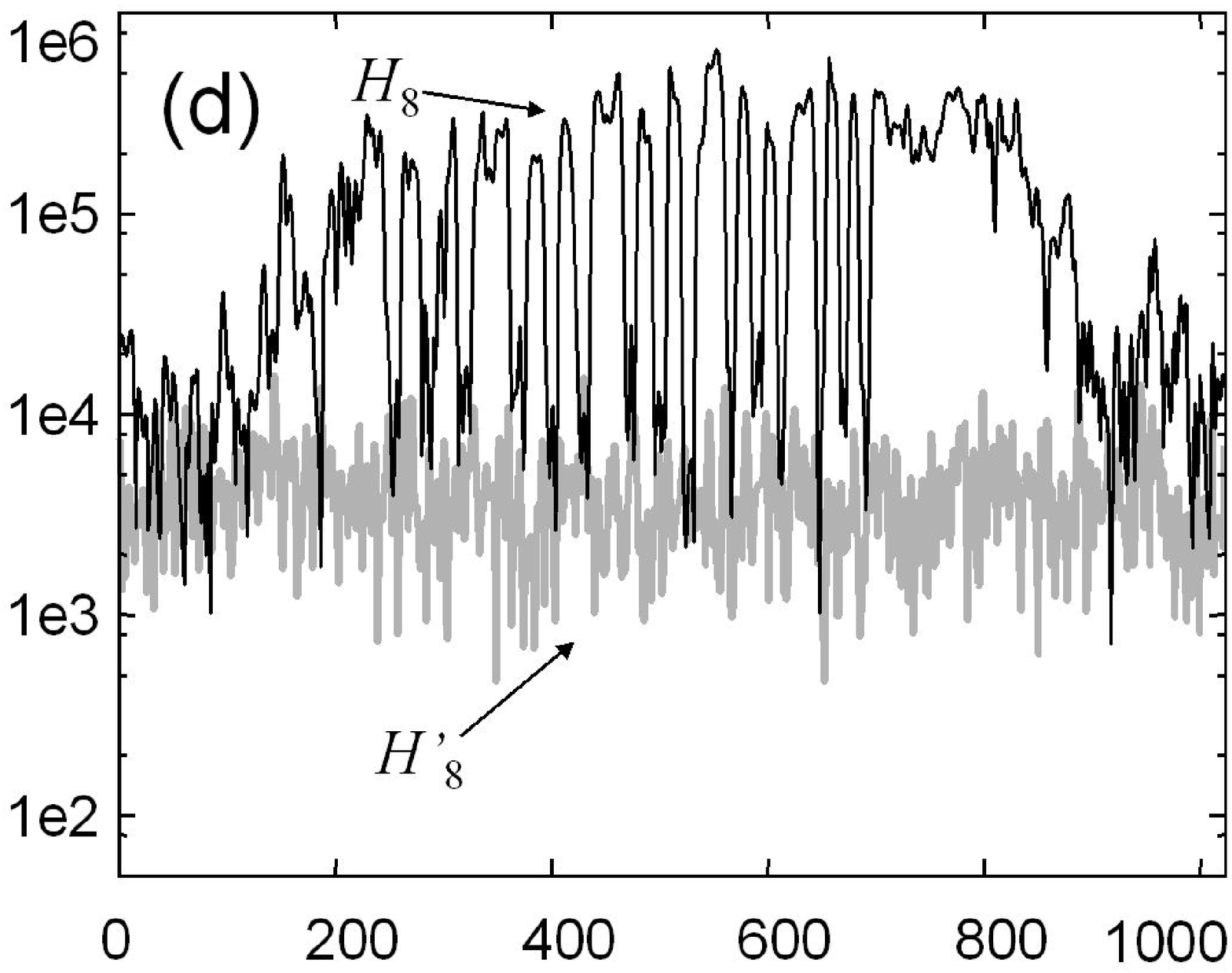}\\
\caption{Same as Fig.\ref{fig_cut_2FFT_282_5} with higher illumination: intensity (i.e. $|H(x,y,D)|^2$) along the  AA' white line of Fig.\ref{fig_images_2FFT_282_4}. The black curves correspond $  H_{1,1}$ (a), $ H_{2}$ (b), $H_{4}$ (c) and $ H_{8}$ (d). The light grey curves in background  correspond  to  $  H'_{2}$ (a), $ H'_{2}$ (b), $H'_{4}$ and $ H'_{8}$ (d). Reconstruction is made with a spatial filter of $282\times 282$ pixels.}
\label{fig_cut_2FFT_282_4}
\end{center}
\end{figure}
In order to perform a more quantitative analysis of these results, we have plotted, in logarithmic scale, the intensity of the reconstructed image signal (i.e. $|H|^2$) along the vertical line AA'.  Out of the  bars, we expect to get a higher signal  $|H|^2$, with some variation, because the laser illumination is not uniform,  and because illumination level and SNR are low. Within the black bars, the signal  $|H|^2$ must go down to zero, and one expects to see grooves corresponding to the black bars. In order to decreases the noise of the curves, we have averaged $|H|^2$ over 10 pixels along the $x$ direction. Since the bars are horizontal and since the width of the bar is much larger than 10 pixels, we expect to keep roughly the same contrast for the grooves than without averaging  (but with less noise). In order to visualize the intrinsic noise of the holographic detection, we also have plotted in light grey the cuts obtained  with the same number of frames recorded without object  illumination (i.e. with odd frames).
Figure \ref{fig_cut_2FFT_282_5} shows the plots  obtained from the Fig.\ref{fig_images_2FFT_282_5}   data, i.e.  from the frames recorded at  very low level of illumination.  Figure \ref{fig_cut_2FFT_282_4} is similar to Fig.\ref{fig_cut_2FFT_282_5} and shows  the plots   obtained from the Fig.\ref{fig_images_2FFT_282_4} data, i.e.  from the frames recorded at  slightly higher illumination.

Plot of Fig. \ref{fig_cut_2FFT_282_5} (a) is obtained with $H_{1,1}$ i.e. with one  frame with illumination,  since $H_{1,1}$ is obtained from the difference of  two frames ($I_0$ and $I_1$).  We have also plotted the cut obtained with  $H'_2$ i.e. with  two frames without object  illumination (i.e. with the same number of frames that for $H_{1,1}$). The holographic signal is extremely low, and the vanishing grooves corresponding to the black bars are not visible. Nevertheless, the curve obtained from $H_{1,1}$ is slightly above the one obtained from  $H'_{2}$.
Plot of Fig. \ref{fig_cut_2FFT_282_5} (b) is obtained with $H_{2}$ i.e. with two frames with illumination. To visualize the noise,  the plot obtained with $H'_2$ is also drawn.   Since the SNR of the image is higher, the  grooves from the black bars  are now visible. Out of the grooves the curve obtained with illumination is "above" the curve obtained without. Within the grooves, the curves obtained with and without object  illumination are roughly at the same level.
Plots that are displayed on Fig. \ref{fig_cut_2FFT_282_5} (c) and (d) are similar to those of Fig. \ref{fig_cut_2FFT_282_5} (b). Out of the grooves the curves obtained with illumination    are ''above'' the curves obtained without, and since the number of frames increases  $|H|^2$ becomes higher and higher. Within the groove  the curves obtained with an without object  illumination are at roughly the same level. A  good estimate of the holographic detection noise can be thus obtained by comparing the curve obtained with illumination, with the curve obtained in the same conditions (same number of images, same spatial filter) without object  illumination.

Plots of Fig. \ref{fig_cut_2FFT_282_4} are similar to the ones of Fig. \ref{fig_cut_2FFT_282_5}, but with  higher illumination.  As expected $|H|^2$ is higher out of the grooves.     Within  the grooves, $|H|^2$ decreases, and the relative contrast of the groove becomes  higher from Fig. \ref{fig_cut_2FFT_282_4}(a) to Fig. \ref{fig_cut_2FFT_282_4}(d). Nevertheless, since the holographic signal $|H|^2$ becomes higher and higher, $|H|^2$ do not reach, within  the groove, the  noise floor that correspond to the data recorded without object  illumination.

\subsection{Average intensity $\overline {|H|^2}$ }

In order to determine how the signal and the noise scale with various   experimental and reconstruction parameters (parity of the frames, number of frames, and area of the spatial filter), we  have averaged the pixel intensity over the whole reconstructed image, and we have plotted, for the even and odd frames, the average intensity $\overline {|H|^2}$ as a function of either the number of frames, or the  area of the spatial filter. The average intensity obtained with illumination is then mainly related to the holographic  signal from the USAF target, while the intensity obtained without object  illumination is  related to the detection noise.
To simplify the analysis, we have considered the case of the higher illumination level of Fig.\ref{fig_images_2FFT_282_4} and Fig.\ref{fig_cut_2FFT_282_4}(a). In that case, the grooves can be seen with only one illuminated frame (i.e. for $H_{1,1}$) as seen on Fig.\ref{fig_cut_2FFT_282_4}. This means that signal from the USAF target overcomes the noise in the worst case.

\begin{figure}[]
\begin{center}
\includegraphics[height =6.5 cm,keepaspectratio=true]{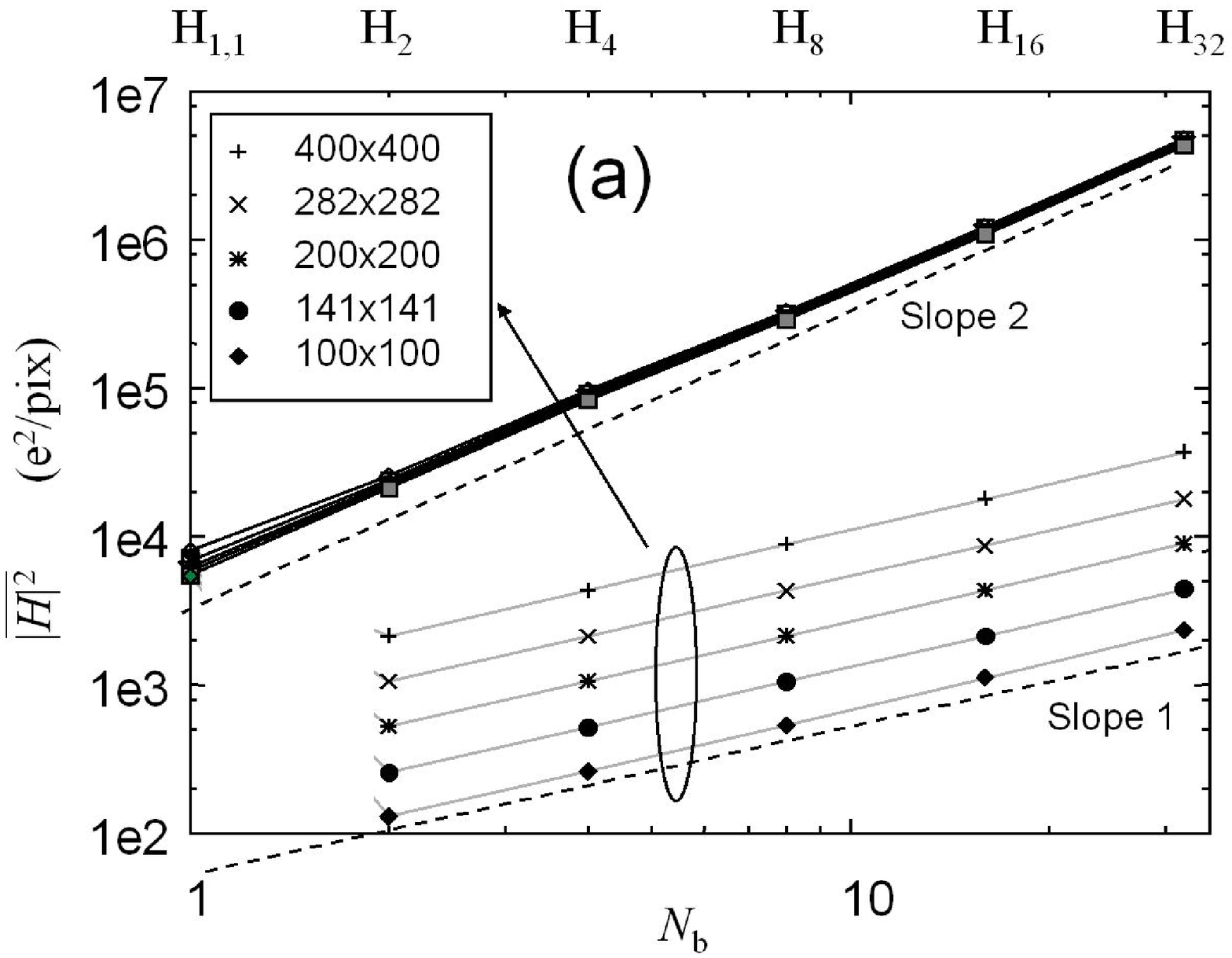}
\includegraphics[height =6.5 cm,keepaspectratio=true]{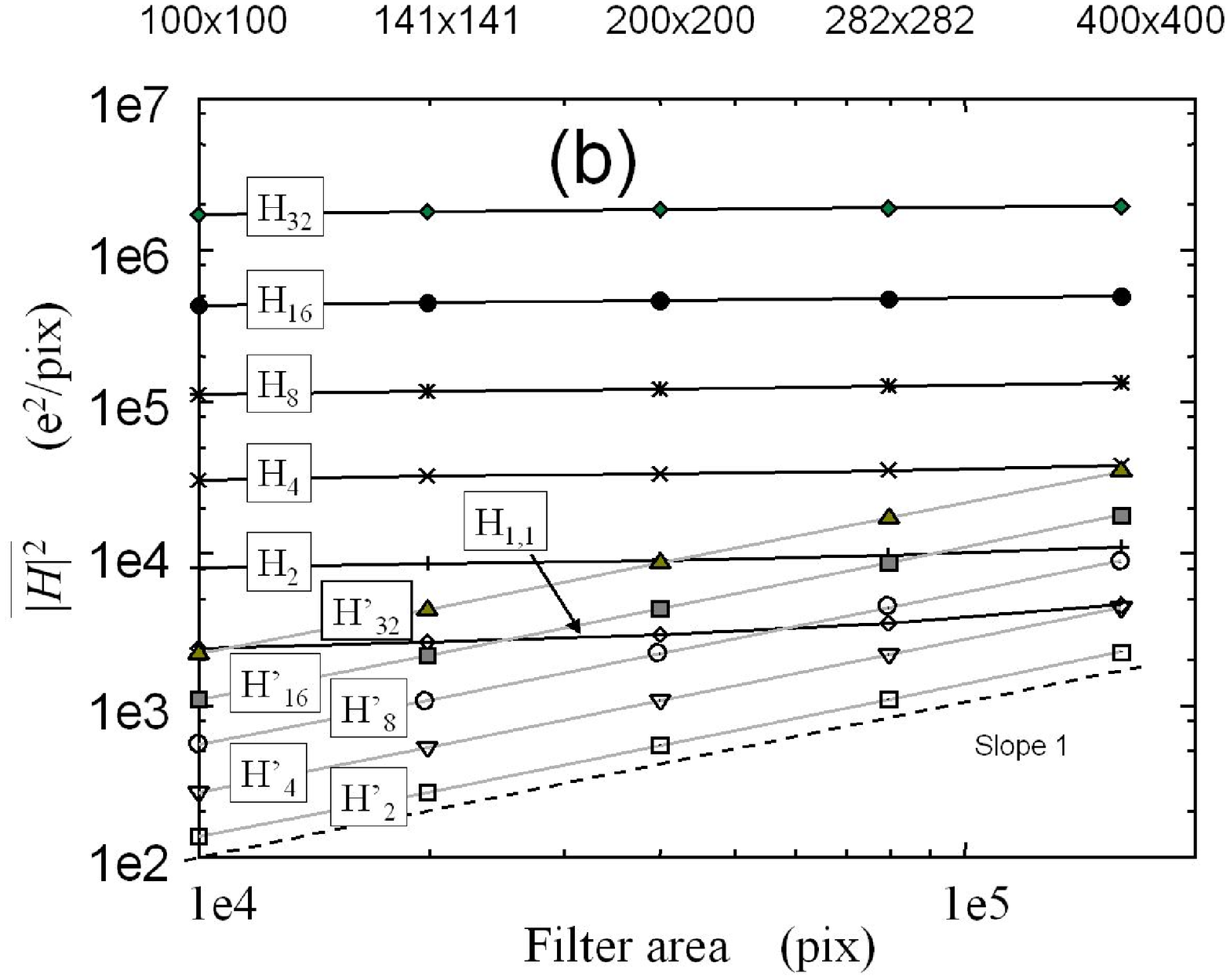}
\caption{Average intensity $\overline{ |H|^2}$: (a)  as a function of the number $N_b $ of frames recorded with illumination: $N_b=1$ for $H_{1,1}$ or $H'_{1,1}$, $N_b=2$ for $H_2$ or $H'_2$,..., and  $N_b=32$ for $H_{32}$ or $H'_{32}$, and  (b) as a function of the area  of the spatial filter  (in pixel units). Even frames curves are black, odd frames curves are light grey.   }.
\label{Fig_curve_intensity_4}
\end{center}
\end{figure}

Figure \ref{Fig_curve_intensity_4} shows the averaged (over the $1024\times 1024$ pixels of the reconstructed images) intensity  $\overline{|H|^2}$ . This intensity is plotted for both even and odd frames as  a function of the number $N_b $ of frames recorded with illumination in Fig. \ref{Fig_curve_intensity_4}(a) or as a function of the area of the spatial filter in Fig. \ref{Fig_curve_intensity_4}(b).

Consider first Fig. \ref{Fig_curve_intensity_4} (a).  For even frames (i.e. with  illumination)    the average intensity   $\overline {|H|^2}$    increases quadratically with the number  of frames (with a Log Log slope close to 2).
This is expected, since we perform a coherent acquisition of the holographic data. The amplitude of the holographic signal (i.e. $H$) increases  linearly with acquisition time, i.e. linearly with $N_b$. The  intensity   $\overline{|H|^2}$ increases thus like $N_b^2$. We must notice that the slope is slightly lower that 2. This may be due  to  loss of coherence of the holographic detection for long time (i.e. large $N_b$) and  to noise for short time (i.e. small $N_b$), since the noise contribution becomes  larger when the detected signal goes down. Note that the noise contribution strongly depends on spatial filter area. As a consequence, the average energy varies quite a bit  for  the first point of the curves ($N_b=1 $ i.e. $H_{1,1}$ ) as seen on Fig. \ref{Fig_curve_intensity_4} (a).

With odd frames (i.e. without  illumination), the average intensity  $\overline{|H|^2}$  increases linearly with  the number  of frames $N_b $  with slope equal to 1, as expected.  Since the phase of the noise changes from one frame to the next,  we perform  an incoherent acquisition of the noise. The amplitude $H$ of the holographic noise  increases thus like the square root of the acquisition time, i.e. like $\sqrt{N_b}$. The  intensity   $\overline{|H|^2}$ increases thus like $N_b$.

\begin{figure}[]
\begin{center}
\includegraphics[width =3.8 cm,keepaspectratio=true]{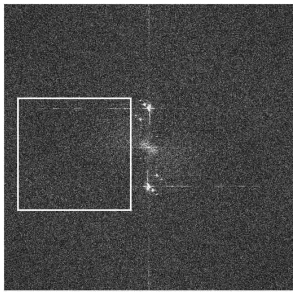}
\caption{Reciprocal space intensity image ${\tilde H}(k_x,k_y,0)$ obtained by FFT from $H'_{2}$ (two odd frames without object  illumination), and typical location of the    spatial filter ($400\times 400$ pixels). }.
\label{Fig_Fig_image_noise_k}
\end{center}
\end{figure}

Figure \ref{Fig_curve_intensity_4} (b) analyses the dependence of the average intensity with the spatial filter area. For even frames (i.e. with  illumination)    the average intensity   $\overline{|H|^2}$ does not strongly depend  on the spatial filter area: the black curves are flat. This is expected since most of the holographic signal  energy is within the low spatial frequency modes.
For odd frames  contrarily (i.e. without  illumination), the average intensity   $\overline{|H|^2}$ varies linearly with the spatial filter area: the light grey curves of  Fig. \ref{Fig_curve_intensity_4} (b)  have a slope equal to 1.
This is expected since we  detect noise. The noise is  incoherent and  homogeneous  in  the spatial frequency region that is selected by the spatial filter. This point is illustrated by Fig. \ref{Fig_Fig_image_noise_k} that shows a reciprocal space intensity image ${\tilde H}(k_x,k_y,0)$ obtained without object  illumination. The noise density is flat within the spatial filter region. Thus, the noise energy varies linearly with the spatial filter area.
%

  %

\begin{figure}[]
\begin{center}
\includegraphics[width =4.0 cm,keepaspectratio=true]{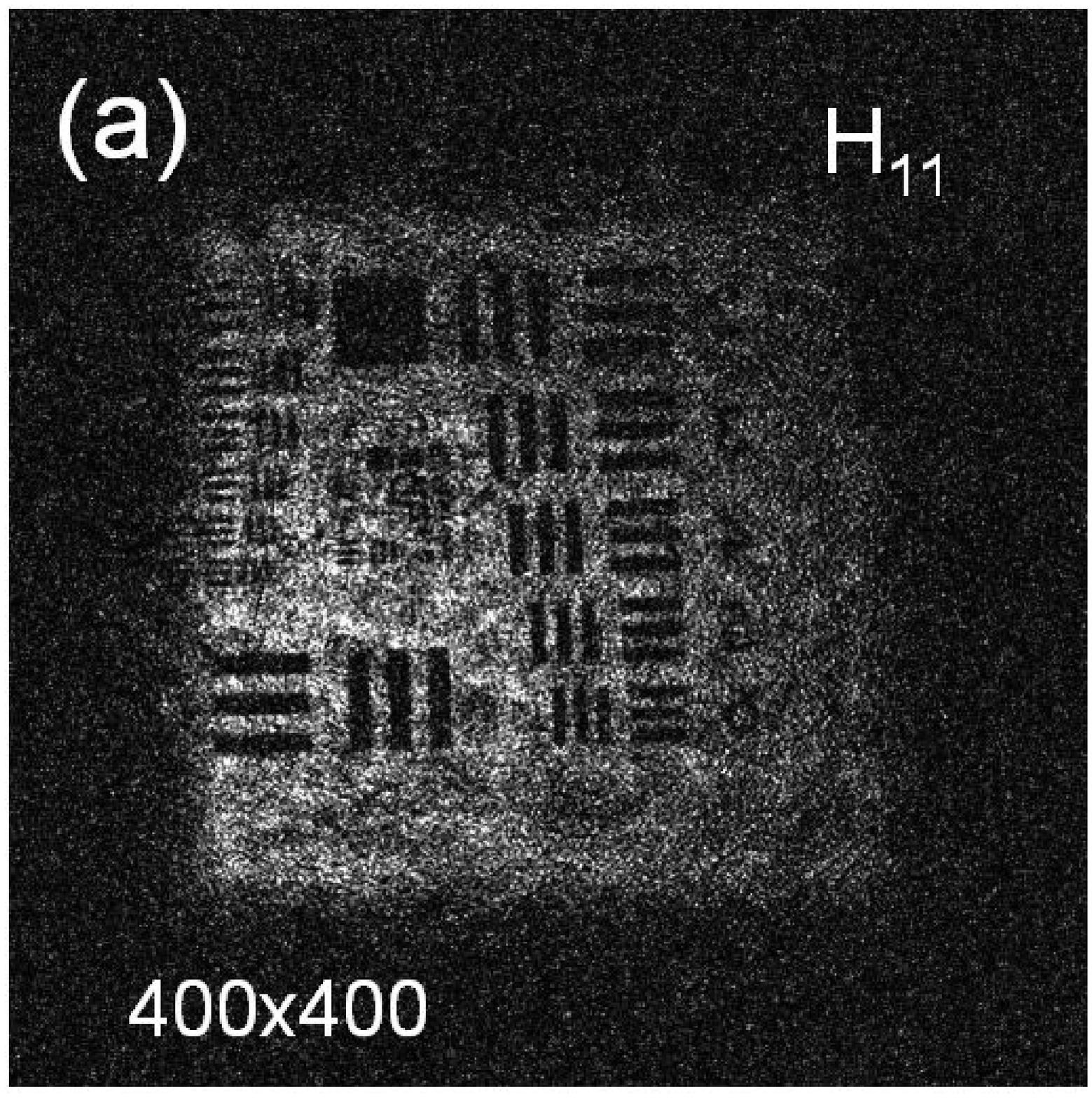}
\includegraphics[width =4.0 cm,keepaspectratio=true]{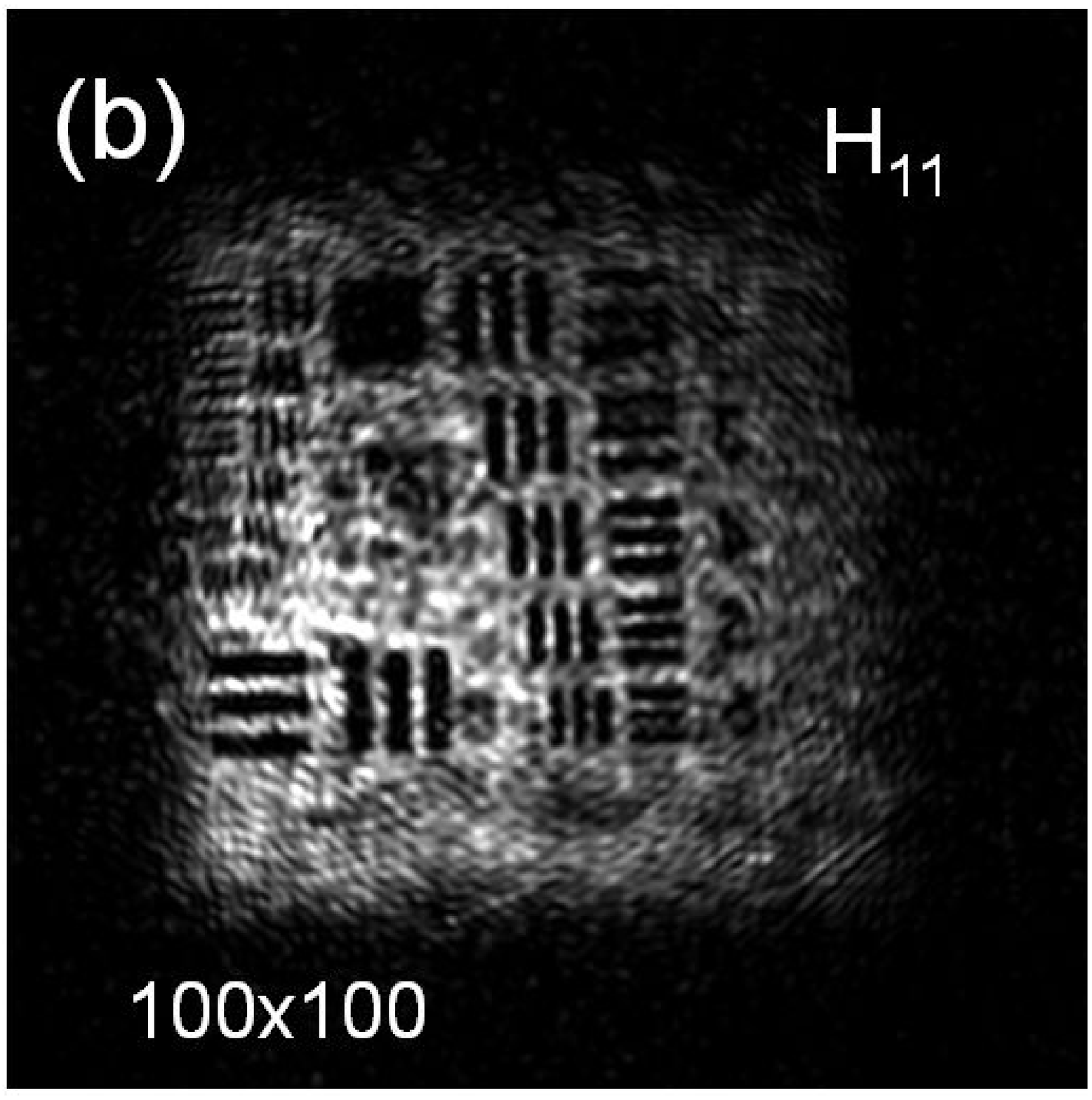}\\
\caption{Reconstructed image from $H_{1,1}$ for a $400\times 400$ pixels (a) and  $100\times 100$ pixels  spatial filter.  }.
\label{Fig_Fig_image_SP}
\end{center}
\end{figure}

Since the signal increases quadratically  with time, while the background noise increases linearly, the signal to background ratio  of the holographic detection increases linearly with time as expected for an ideal coherent detection. This mean  that the equivalent noise is a fixed amount of energy that do not depends on time.
On the other hand, since the signal is roughly flat with the area of the spatial filter, while the noise increases linearly with it, the signal to background ratio  is  proportional to the inverse of the spatial filter area.

One must nevertheless avoid to reduce too much the filter area, since  this area affects  the image quality.  This  point is illustrated by Fig. \ref{Fig_Fig_image_SP} that compares reconstructed images of the USAF target
made from $H_{1,1}$ with the slightly higher illumination level of Fig. \ref{fig_images_2FFT_282_4}. The spatial filter is $400\times 400$ pixels for Fig.\ref{Fig_Fig_image_SP}(a) and $100\times 100$ pixels for Fig.\ref{Fig_Fig_image_SP}(a). Since  the area is larger on Fig. \ref{Fig_Fig_image_SP}(a),  the reconstruction involves  more spatial frequency components yielding more noise, but also better  resolution as seen by comparing Figs. \ref{Fig_Fig_image_SP} (a) and (b).

\subsection{Analysis of the noise}

As we have considered the case of an object (the USAF  target) weakly illuminated, noise is not related to the signal of the object. It can therefore be measured in the absence of illumination of the object. This is what we have confirmed by the analysis of curves of Fig. \ref {fig_cut_2FFT_282_4}, which shows that the intensity $|H|^2 $, obtained with illumination in the black parts of the reconstructed image (bottom of the grooves associated with the image of the horizontal bars of the USAF target) is the same as obtained without object  illumination.

Without object  illumination, the camera records only the attenuated signal from the laser (i.e. the LO reference beam), and the observed noise is equal to the sum of the technical noises with  shot noise. The technical noises are mainly the  electronic noise of the camera, the quantization noise of the camera ADC (Analog to Digital Converter) and the laser intensity noise. Shot noise results from different stochastic processes (laser emission, absorption by optical attenuators, photo conversion within the pixels of the camera) that yield an electronic signal corresponding for each frame and each pixel to an integer number of photo electron. This number is affected  by a random Poisson noise (with a standard deviation equal to the root of the number of photo electrons).

Let us consider the  signal $I_n(x,y)$ detected on frame $n$ and pixel $(x,y)$. Because of shot noise, $I_n(x,y)$ is the sum of a statistical average component $\langle I(x,y) \rangle$ with a shot noise component $o(x,y,n)$, which is random  for each pixel $(x,y)$ and each frame $n$. Thus we have:
\begin{eqnarray}\label{Eq_noise o1}
  I_n(x,y) =  \langle I(x,y) \rangle + o_n(x,y)
\end{eqnarray}
Since $I_n(x,y)$  is measured in photo electron  units (e), the fluctuations $o_n(x,y)$ obey Poisson statistic (or Gaussian since $I_n(x,y) \gg 1 $) with :
\begin{eqnarray}\label{Eq_noise o2}
 \langle o_n(x,y) \rangle &=& 0\\
\nonumber  \langle |o_n(x,y)|^2 \rangle  &=& \langle I(x,y) \rangle
\end{eqnarray}
Note here that  Eq. (\ref{Eq_noise o2})  that defines $o_n(x,y) $  is only valid  if  $ I_n(x,y) $ is expressed in photoelectron  units (e). Fluctuations $o_n(x,y) $ therefore depend on the gain of $ G $ the camera. The statistical average $\langle I_n(x,y) \rangle$, and the shot noise $o_n(x,y)$ terms of Eq. (\ref{Eq_noise o1}) are  not accessible to measurement, and cannot be separated. Therefore, one cannot remove the shot noise term from a measured frame $I_n(x,y)$.

In the case of a camera, the photo electronic signal corresponds to a relatively small number of photoelectrons e, and shot noise is relatively large. In our experiment for example, the local oscillator power is adjusted so that $ I_n $ corresponds to half saturation of the CCD camera. Since our camera is 12 bits, $ I_n $ is thus  $I_n \sim 2000$ DC  (Digital Count), i.e. $I_n \sim 10^4$ e   (since  $ G = $ 4.41 e / DC). The shot noise  $o_n(x,y) \sim 100$ e is thus much larger than   electronic (30 e in our case) and   quantization  ($4.4$ e) noise.

The relative importance of shot noise with respect to the technical noises  depends heavily on the experimental situation. In our case, the signal detected by the camera is dominated by the LO, and the signal is strong. The shot noise is important, and dominates the camera noise. In the case of a direct imaging (without interference with a LO) of an object weakly illuminated, the photoelectric signal is much weaker, and the electronic noise of the camera can become the dominant noise.

\begin{figure}[]
\begin{center}
\includegraphics[width =6 cm,keepaspectratio=true]{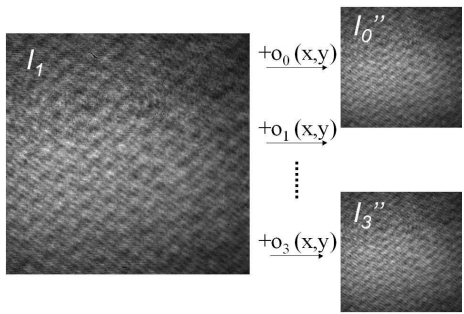}\\
\caption{Principe of calculation of the  Monte Carlo odd frames $I''_n(x,y)$ from $I_1(x,y)$ recorded without object  illumination.  }.
\label{Fig_fig_monte_carlo}
\end{center}
\end{figure}

To check if our holographic noise detection is or is not limited by shot noise, we have calculated the odd frames by Monte Carlo simulation, and we have calculated the corresponding noise. The main advantages of Monte Carlo simulation are that it is easy to implement and robust. Moreover, since the photons are emitted in a discrete and random way, Monte Carlo calculation yields intuitive description of our experiment, and of its noise.

In our case, the detected signal is large: $I_n(x,y)\gg 1$. The statistical average $\langle I(x,y) \rangle$ is much larger than the fluctuation $o_n(x,y)$. It  can be thus replaced in Eq. (\ref{Eq_noise o2}) by the signal $I_n(x,y)$ recorded on the same frame, or, since this signal is roughly the same for all frames, by the signal $I_1(x,y)$ recorded without object  illumination on frame 1. We have:
\begin{eqnarray}\label{Eq_noise o3}
 \langle o_n(x,y) \rangle &=& 0\\
\nonumber  \langle |o_n(x,y)|^2 \rangle &\simeq &    I_1(x,y)
\end{eqnarray}
One can calculate $o_n(x,y)$ by Monte Carlo. Since all the holograms are built by making differences of frames (see Eq.\ref{Eq_holograms_H} and Eq.\ref{Eq_holograms_H11}), the contribution of the  statistical average cancels in the holograms. As illustrated by Fig. \ref{Fig_fig_monte_carlo}, one can thus  replace $\langle I(x,y) \rangle$ by  $I_1(x,y)$ in Eq.\ref{Eq_noise o1} to  calculate Monte Carlo odd frames $ I''_n(x,y) $, and Monte Carlo odd holograms $H''_x$
\begin{eqnarray}\label{Eq_monte_carlo}
 I''_{n}(x,y) &=& I_{1}(x,y) +o_{n}(x,y)
\end{eqnarray}
The Monte Carlo odd holograms  $H''_{n}(x,y)$ are  similar to the measured odd holograms  $H'_{n}(x,y)$.  They are calculated by replacing $I_{2n}(x,y)$ by $I''_{2n+1}(x,y)$ in Eq. \ref{Eq_holograms_H}.

\begin{figure}[]
\begin{center}
\includegraphics[width =3.5 cm,keepaspectratio=true]{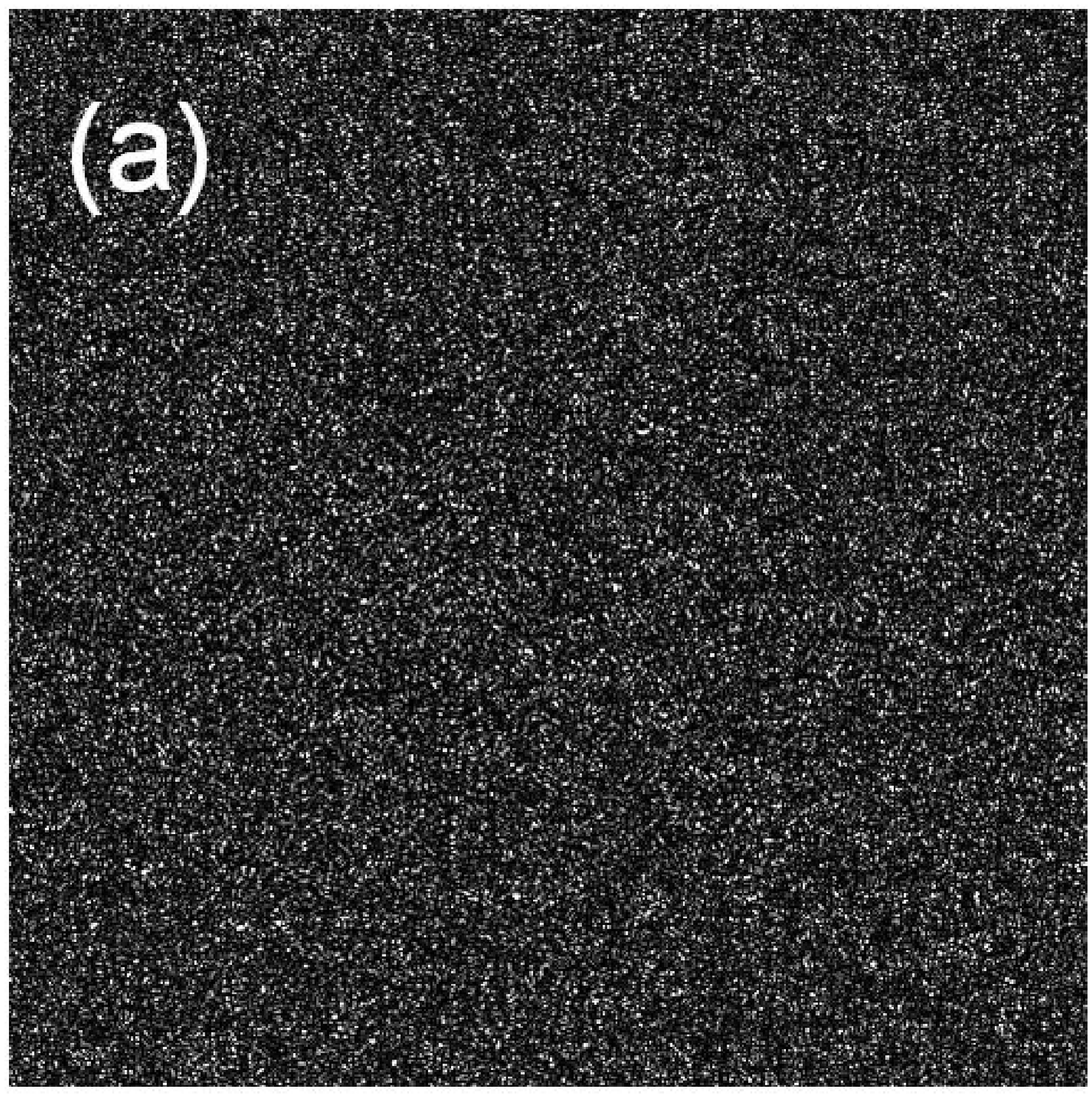}
\includegraphics[width =3.5 cm,keepaspectratio=true]{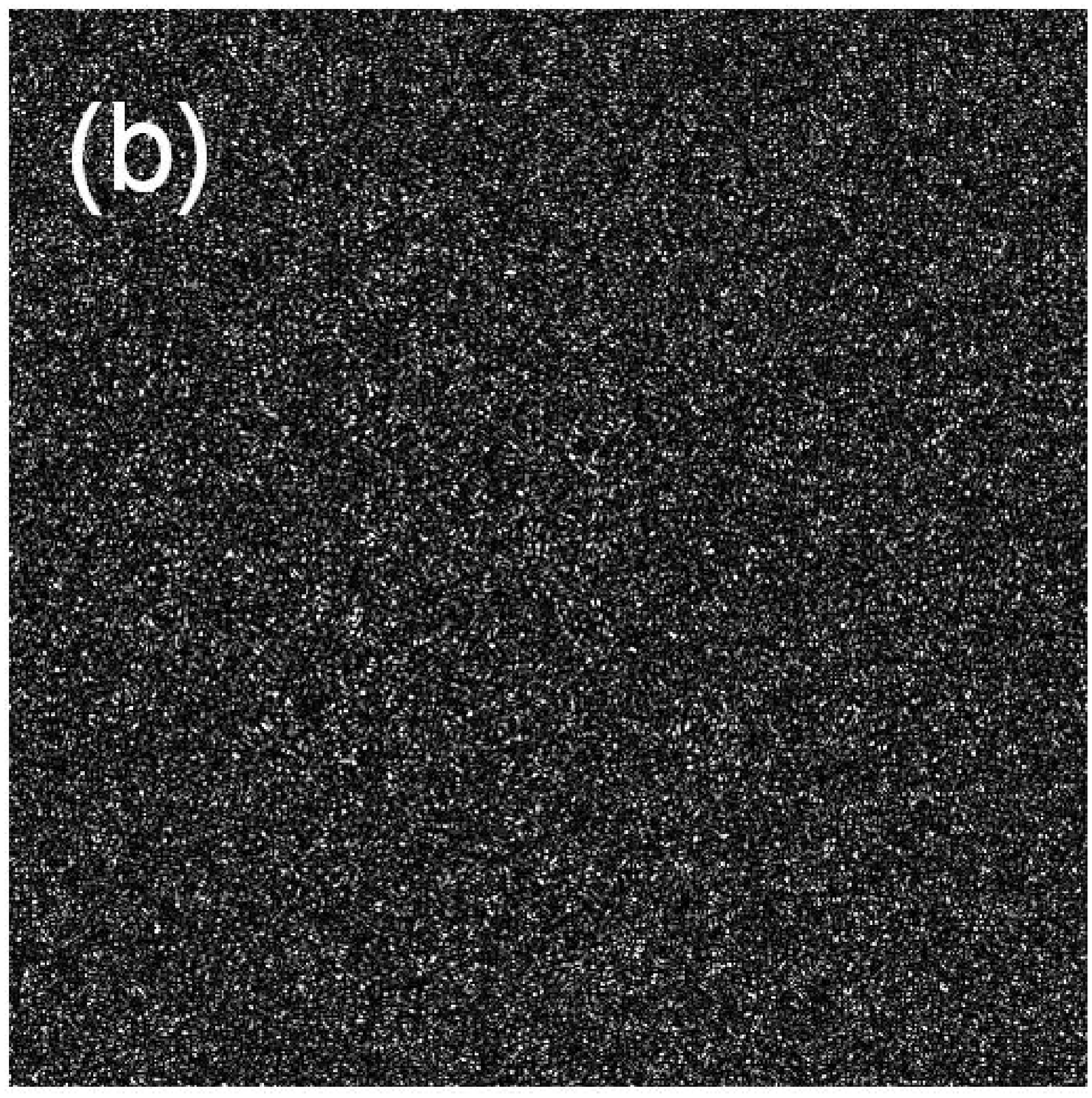}\\
\includegraphics[width =4.2 cm,keepaspectratio=true]{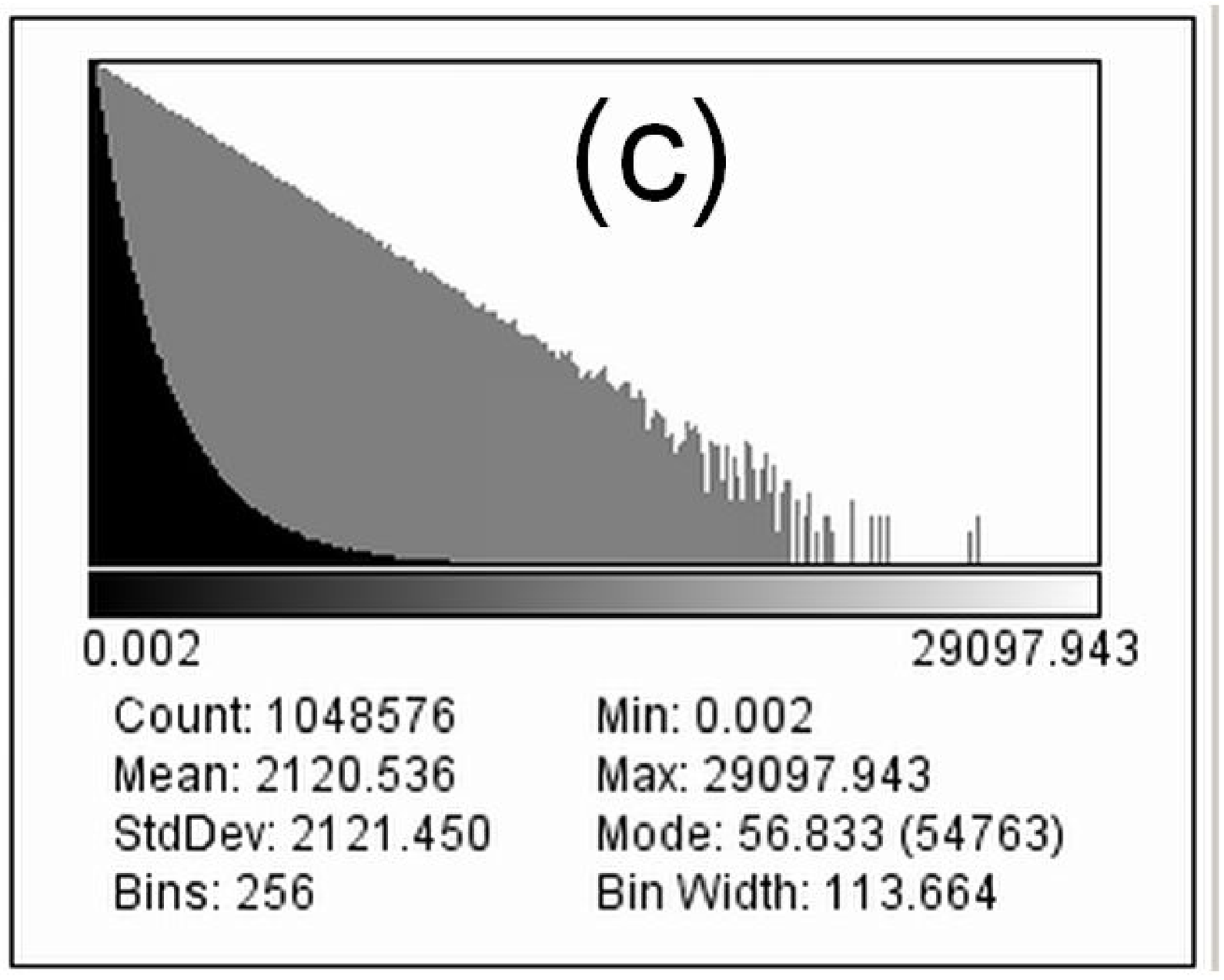}
\includegraphics[width =4.2 cm,keepaspectratio=true]{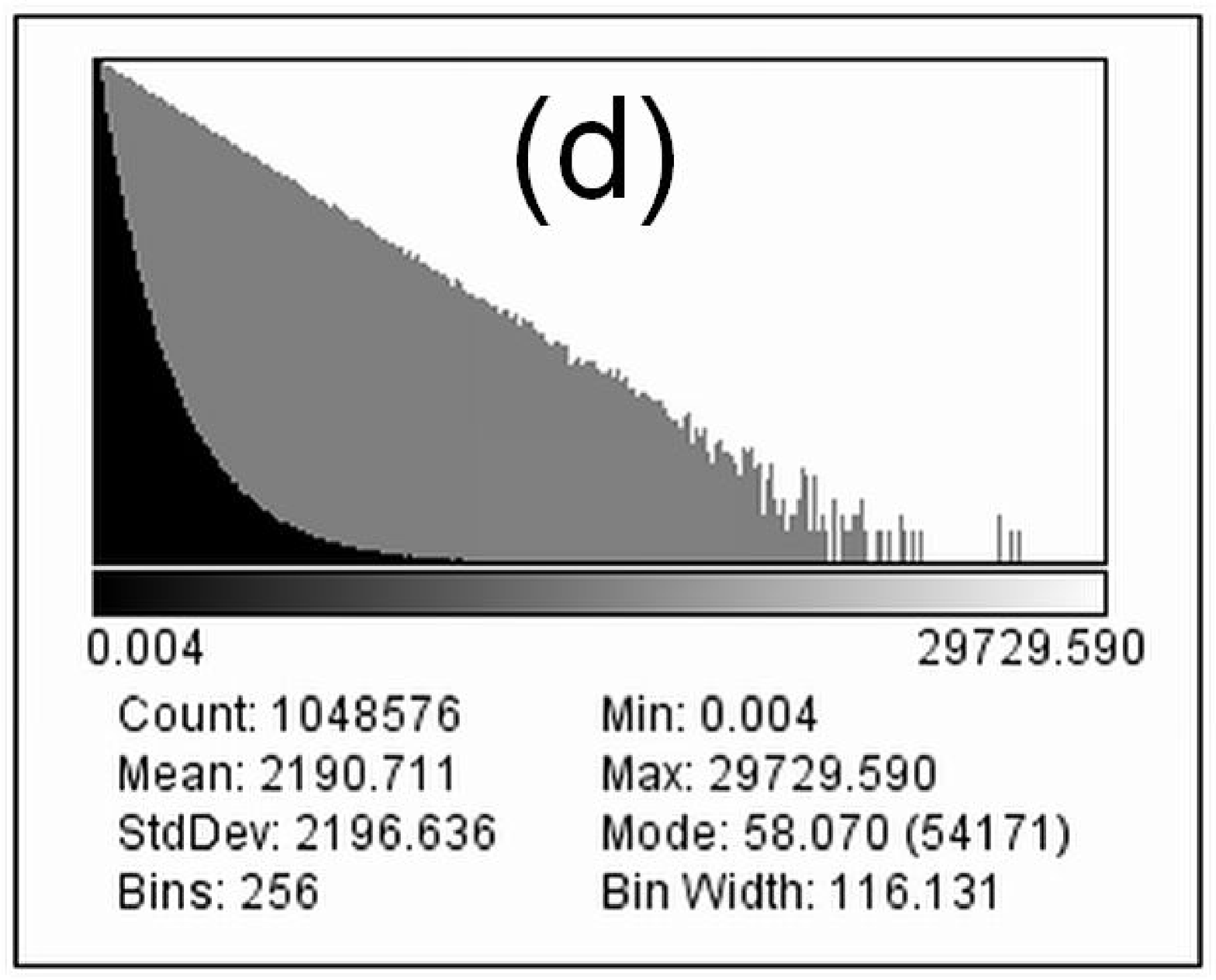}
\caption{Reconstructed images calculated from 4 frames recorded  without object  illumination i.e. from  $H'_4$ (a), and from 4 Monte Carlo frames  i.e. from $H''_4$ (b). Histogram of the intensity $|H|^2$ displayed in linear (black) and logarithmic scale (light grey): without object  illumination (c), and with Monte Carlo   (d).
The reconstruction is made with a $282 \times 282$ pixels spatial filter.  }
\label{Fig_comp_noise}
\end{center}
\end{figure}
\begin{figure}[]
\begin{center}
\includegraphics[width =6.5 cm,keepaspectratio=true]{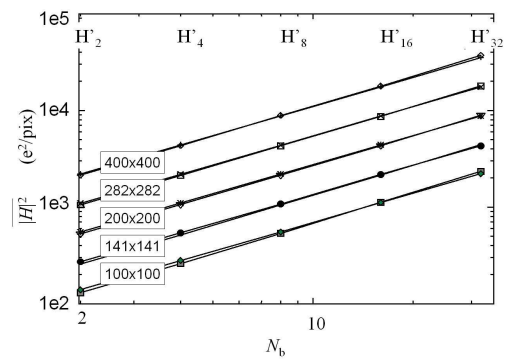}\\
\caption{Average intensity $\overline{|H|^2}$ as a function of the number $N_b $ of frames: $N_b=2$ for $H'_2$ or $H''_2$,  $N_b=4$ for $H'_4$ or $H''_4$,..., and  $N_b=32$ for $H'_{32}$ or $H''_{32}$. The frames are either recorded without object  illumination $H'_n$ or simulated by Monte carlo $H''_n$. The images have been reconstructed with 2FFT. The size of the spatial filter is $400\times 400$, $282\times 282$,  $200\times 200$, $141\times 141$ and $100\times 100$. }
\label{Fig_curve_noise_4}
\end{center}
\end{figure}

Figure  \ref{Fig_comp_noise}  compares the reconstructed images from 4 frames holograms  measured without illumination of the target (i.e. calculated from $H'_4  $) in Fig.\ref{Fig_comp_noise}(a), with the Monte Carlo ones (i.e. calculated from $H''_4  $) on Fig. \ref{Fig_comp_noise}(b).
The two images are very similar and show a Gaussian distribution for the holographic complex noise amplitude $ H (x, y, D) $, and an exponential  distribution for the intensity $ | H (x, y, D) | ^ 2 $. Using ImageJ, we have plotted the  histogram of intensity $ | H (x, y, D) | ^ 2 $ for the measured (Fig.\ref{Fig_comp_noise}(c)) and the  Monte Carlo reconstructed images (Fig.\ref{Fig_comp_noise}(d)).
The histograms plotted in logarithmic scale show that the measured and  Monte Carlo distributions are exponential. This result is confirmed by the average intensity (measured: 2120;  Monte Carlo: 2190) which is equal to the standard deviation (measured: 2121;  Monte Carlo: 2196).
We can also note that average intensities and standard deviations are the same for the measured and  Monte Carlo data. The noise observed in experiments is thus  equal to the  shot noise that can be calculated by Monte Carlo knowing the gain of $ G $ the camera.

To confirm this results we have calculated the average intensities measured and simulated for holograms  with different number of frames: $ N_b = 2 $ to 32, and for different areas of the spatial filter: $ 100\times 100$  to  $400\times 400$ pixels. The results are presented in  Fig.\ref{Fig_curve_noise_4}.  Note that the measured points have already been  plotted in Fig.\ref{Fig_curve_intensity_4} (a) (light gray curves).
Figure \ref{Fig_curve_noise_4} shows that,  the measured average intensities  are extremely close  to the Monte Carlo simulated ones  for all numbers of frames, and all spatial filter areas.  The noise observed in experiment is thus dominated by  shot noise.

It should be noted that thanks to the heterodyne gain, our setup is immune to electronic noises. Similar results are expected with any well-built scientific camera that uses low noise electronic in particular for the amplifying stage before analog to digital conversion. Industrial cameras that may exhibit electronic noise must be avoided. It should also be noted that the CCD exposure time (for 1 frame) does not play any role, since it do not appear in the Monte Carlo simulation. Results depend only on the number of photo electrons. Thus, the exposure time plays no rule as far as the dark noise of the camera remains lower than the local oscillator signal (LO).

The scaling laws observed for  the experimental noise can be interpreted physically.
The shot noise is a white noise, which is random and Gaussian in spatial and temporal frequencies. The holographic detection is coherent  in space and time. As a result, on one hand (space),  the detected noise is proportional to the number of spatial modes that are detected, i.e. to  the number of pixels of the spatial filter.
On the other hand (time), the noise is  proportional to the product of the measurement time $ T $ by the detection bandwidth, which is here equal to $ 1/T $. As a result, the equivalent noise does not depend on time, i.e. on the number of frames $N_b$.

One must notice that the  holographic signal $H(x,y,0)$, detected by the camera, is the interference of the optical field scattered by the object $E$ with the reference beam local oscillator field $E_{LO}$. It corresponds to $E_{LO}^* E$. This signal  is detected and integrated over time by the camera yielding a measurement in photoelectron units (e). This means that the reconstructed image, which corresponds  to $|H(x,y,D)|^2$ (i.e. to $|E_{LO}^*|^2 \times  |E|^2$ integrated over the measurement times), is measured in  e$^2$ units for each pixel. The vertical axis $|H|^2$ or $\overline {|H|^2}$ of Fig. \ref{fig_cut_2FFT_282_5}, Fig. \ref{fig_cut_2FFT_282_4},  Fig.\ref{Fig_curve_intensity_4} and Fig.\ref{Fig_curve_noise_4} corresponds to the product of the energy of the signal field $|E|^2$ in e units, by the LO field $|E_{LO}|^2$ in e units too. It is then possible to get an absolute calibration of the detected signal.

Consider for  example  the image reconstructed from $H'_4$ with a $282\times 282$ spatial filter displayed on Fig.\ref{Fig_comp_noise} (a). In all our experiments, the  frame signal $I_n$  averaged over the pixels is $\overline{I_n} \simeq 6800$ e. The LO signal is thus $|E_{LO}|^2$=6800 e for one frame, and 27200 e for 4 frames.
Since  $\overline{|H|^2}$ is 2120.53 e$^2$ in that case (see Fig.\ref{Fig_comp_noise} (c) ), the equivalent noise $|E|^2$ is here $2120/27200=  0.0779$ e. This result is obtained for a $282\times 282$ spatial filter. It corresponds to an equivalent noise of $0.0779\times (1024/282)^2= 1.02 \simeq 1$ e without spatial filtering.  This result is general and does not depends on the number of frames, nor on the spatial filter area. The equivalent noise corresponds thus 1 photoelectron per pixel of the spatial filter whatever the number of frames or  spatial filter area is  \cite{bachor_1998,verpillat2010digital}.

It should be noted that the experimental and Monte Carlo simulated signals depends on the gain of the camera $G$ with different factors. On one hand, the frame signal $I_n(x,y)$, and the measured holograms $H'_n$  are expressed in photo electron units. They are proportional to $G$. On the other hand, the Monte Carlo simulation fluctuations $o_n(x,y)$ are calculated from Eq. \ref{Eq_noise o3}. They depend on $G$ through $ I_1(x,y)$. The fluctuations  $o_n(x,y)$ varies thus like the  square root of $G$. As result, the Monte Carlo holograms $H''_n(x,y)$ that are calculated from the Monte Carlo frames $I''_n(x,y)$ (defined by Eq. \ref{Eq_monte_carlo}), are proportional to $\sqrt{G}$.

Thus, the agreement of the measured noise with the Monte Carlo calculated shot noise depends on the exact value of the camera gain $G$. This  gain $ G $   has been calibrated  by the camera manufacturer (here $G =$ 4.41 e/DC) by illuminating the camera  with a very clean white light source, which must be stable   spatially and temporally and by  assuming that the observed fluctuations of the camera signal in time or in space result from  shot noise \cite{Newberry1998}.
The holographic measurement we have done without object illumination is thus very close to what was done to calibrate the camera. Nevertheless, our light source is much less clean. Our laser may exhibit  temporal fluctuations, and, because of stray reflections, the power of the LO beam is not constant on the camera detecting area, as shown on image of  frame $ I_1 $ at left in Fig.\ref{Fig_fig_monte_carlo}.
The good agreement between the measured noise and the noise calculated by Monte Carlo from the  gain $G$ of the camera, shows that the fluctuation we got with a laser are the same than  the fluctuation observed by the camera manufacturer with a white light source yielding the same average camera signal.
This means that the defects of our experimental setup (temporal fluctuations of the laser intensity and spatial variations of the LO signal) do not seem to bring excess of noise with respect to shot noise.

We think we  are insensitive to  defaults of our setup  because we do both heterodyne and off-axis holography. The holographic signal is thus modulated both in time (heterodyne) and space (off-axis). The unwanted low frequency time components are filtered off by building  holograms with  many  frames (see Eq. \ref{Eq_holograms_H} and Eq. \ref{Eq_holograms_H11}), while  the low frequency space components are filtered off  by  spatial filtering.

\section{Conclusion}

By recording hologram of a  USAF target under weak illumination with our off axis heterodyne holography setup,  we have found that the signal and noise scales as expected with a shot noise limited experiment. The signal for the reconstructed image intensity $|H|^2$ is proportional to the square of the time, while the noise, which is equal to the expected shot noise \cite{verpillat2010digital},  is proportional to the time, and to the number of modes involved in the holographic reconstruction (area of the spatial filter in pixels units). These results validate previous weak holographic signal  experiments \cite{gross2003shot,gross2007digital,gross2005heterodyne,warnasooriya2010imaging,verpillat2011dark,absil2010photothermal,psota2012comparison}

\bibliographystyle{osajnl}

\begin{thebibliography}{10}
\newcommand{\enquote}[1]{``#1''}
\expandafter\ifx\csname url\endcsname\relax
  \def\url#1{\texttt{#1}}\fi
\expandafter\ifx\csname urlprefix\endcsname\relax\def\urlprefix{URL }\fi
\providecommand{\eprint}[2][]{\url{#2}}

\bibitem{Gabor49}
D.~Gabor, \enquote{Microscopy by Reconstructed Wave-Fronts,} Proc. Royal Soc.
  London A \textbf{197}, 454--487 (1949).

\bibitem{macovski1971considerations}
A.~Macovski, \enquote{Considerations of television holography,} J. Mod. Opt. \textbf{18}(1), 31--39 (1971).

\bibitem{goodman1967digital}
J.~Goodman and R.~Lawrence, \enquote{Digital image formation from
  electronically detected holograms,} Appl. Phys. Lett. \textbf{11}(3),
  77--79 (1967).

\bibitem{schnars1994direct}
U.~Schnars, \enquote{Direct phase determination in hologram interferometry with
  use of digitally recorded holograms,} J. Opt. Soc. Am. A \textbf{11}(7), 2011--2015
  (1994).

\bibitem{leith1965microscopy}
E.~Leith, J.~Upatnieks, and K.~Haines, \enquote{Microscopy by wavefront
  reconstruction,} J. Opt. Soc. Am. A \textbf{55}(8), 981--986 (1965).

\bibitem{Schnars_Juptner_94}
U.~Schnars and W.~J{\"u}ptner, \enquote{Direct recording of holograms by a
  {CCD} target and numerical reconstruction,} Appl. Opt. \textbf{33}(2),
  179--181 (1994).

\bibitem{kreis1998principles}
T.~Kreis, W.~Jueptner, and J.~Geldmacher, \enquote{Principles of digital
  holographic interferometry,} in \emph{Proceedings of SPIE}, vol. 3478, p.~45
  (1998).

\bibitem{Yamaguchi1997}
I.~Yamaguchi and T.~Zhang, \enquote{Phase-Shifting digital holography,} Opt.
  Lett. \textbf{18}(1), 31 (1997).

\bibitem{doval_2000}
A.-F. Doval, \enquote{A systematic approach to TV holography.}
  Meas. Sci. Technol. \textbf{11}, R1 (2000).

\bibitem{Schnars_2002}
U.~Schnars and W.~J{\"u}ptner, \enquote{Digital recording and numerical reconstruction
  of holograms,} Meas. Sci. Technol. \textbf{13}, R85 (2002).

\bibitem{pu2004intrinsic}
Y.~Pu and H.~Meng, \enquote{Intrinsic speckle noise in off-axis particle
  holography,} J. Opt. Soc. Am. A \textbf{21}(7), 1221--1230 (2004).

\bibitem{colomb2002polarization}
T.~Colomb, P.~Dahlgren, D.~Beghuin, E.~Cuche, P.~Marquet, and C.~Depeursinge,
  \enquote{Polarization imaging by use of digital holography,} Appl. Opt.
  \textbf{41}(1), 27--37 (2002).

\bibitem{cuche1999digital}
E.~Cuche, F.~Bevilacqua, and C.~Depeursinge, \enquote{Digital holography for
  quantitative phase-contrast imaging,} Opt. Lett. \textbf{24}(5), 291--293
  (1999).

\bibitem{Massig_2002}
J.~H. Massig, \enquote{Digital off-axis holography with a synthetic aperture,}
  Opt. Lett. \textbf{27}, 2179--2181 (2005).

\bibitem{ansari2001elimination}
Z.~Ansari, Y.~Gu, M.~Tziraki, R.~Jones, P.~French, D.~Nolte, and M.~Melloch,
  \enquote{Elimination of beam walk-off in low-coherence off-axis
  photorefractive holography,} Opt. Lett. \textbf{26}(6), 334--336 (2001).

\bibitem{massatsch2005time}
P.~Massatsch, F.~Charri\`{e}re, E.~Cuche, P.~Marquet, and C.~Depeursinge,
  \enquote{Time-domain optical coherence tomography with digital holographic
  microscopy,} Appl. Opt. \textbf{44}(10), 1806--1812 (2005).

\bibitem{Marquet_2005}
P.~Marquet, B.~Rappaz, P.~J. Magistretti, E.~Cuche, Y.~Emery, T.~Colomb, and
  C.~Depeursinge, \enquote{Digital holographic microscopy: a noninvasive
  contrast imaging technique allowing quantitative visualization of living
  cells with subwavelength axial accuracy,} Opt. Lett. \textbf{30}, 468--470
  (2005).

\bibitem{atlan2008heterodyne}
M.~Atlan, M.~Gross, P.~Desbiolles, {\'E}.~Absil, G.~Tessier, and
  M.~Coppey-Moisan, \enquote{Heterodyne holographic microscopy of gold
  particles,} Opt. Lett. \textbf{33}(5), 500--502 (2008).

\bibitem{zhang1998three}
T.~Zhang and I.~Yamaguchi, \enquote{Three-dimensional microscopy with
  phase-shifting digital holography,} Opt. Lett. \textbf{23}(15),
  1221--1223 (1998).

\bibitem{nomura2007polarization}
T.~Nomura, B.~Javidi, S.~Murata, E.~Nitanai, and T.~Numata,
  \enquote{Polarization imaging of a 3D object by use of on-axis phase-shifting
  digital holography,} Opt. Lett. \textbf{32}(5), 481--483 (2007).

\bibitem{yamaguchi2002phase}
I.~Yamaguchi, T.~Matsumura, and J.~Kato, \enquote{Phase-shifting color digital
  holography,} Opt. Lett. \textbf{27}(13), 1108--1110 (2002).

\bibitem{Leclerc2001}
F.~LeClerc, L.~Collot, and M.~Gross, \enquote{Synthetic-aperture experiment in
  visible with on-axis digital heterodyne holography,} Opt. Lett.
  \textbf{26} (2001).

\bibitem{tamano2006phase}
S.~Tamano, Y.~Hayasaki, and N.~Nishida, \enquote{Phase-shifting digital
  holography with a low-coherence light source for reconstruction of a digital
  relief object hidden behind a light-scattering medium,} Appl. Opt.
  \textbf{45}(5), 953--959 (2006).

\bibitem{yamaguchi2006surface}
I.~Yamaguchi, T.~Ida, M.~Yokota, and K.~Yamashita, \enquote{Surface shape
  measurement by phase-shifting digital holography with a wavelength shift,}
  Appl. Opt. \textbf{45}(29), 7610--7616 (2006).

\bibitem{Zhang_1998}
T.~Zhang and I.~Yamaguchi, \enquote{Three-dimensional microscopy with
  phase-shifting digital holography,} Opt. Lett. \textbf{23}, 1221--1223
  (1998).

\bibitem{yamaguchi2001image}
I.~Yamaguchi, J.~Kato, S.~Ohta, and J.~Mizuno, \enquote{Image formation in
  phase-shifting digital holography and applications to microscopy,} Appl.
  Opt. \textbf{40}(34), 6177--6186 (2001).

\bibitem{Leclerc2000}
F.~LeClerc, L.~Collot, and M.~Gross, \enquote{Numerical Heterodyne Holography
  Using 2D Photo-Detector Arrays,} Opt. Lett. \textbf{25}, 716 (2000).

\bibitem{gross2007digital}
M.~Gross and M.~Atlan, \enquote{Digital holography with ultimate sensitivity,}
  Opt. Lett. \textbf{32}(8), 909--911 (2007).

\bibitem{gross2005heterodyne}
M.~Gross, P.~Goy, B.~Forget, M.~Atlan, F.~Ramaz, A.~Boccara, and A.~Dunn,
  \enquote{Heterodyne detection of multiply scattered monochromatic light with
  a multipixel detector,} Opt. Lett. \textbf{30}(11), 1357--1359 (2005).

\bibitem{warnasooriya2010imaging}
N.~Warnasooriya, F.~Joud, P.~Bun, G.~Tessier, M.~Coppey-Moisan, P.~Desbiolles,
  M.~Atlan, M.~Abboud, and M.~Gross, \enquote{Imaging gold nanoparticles in
  living cell environments using heterodyne digital holographic microscopy,}
  Opt. Express \textbf{18}(4), 3264--3273 (2010).

\bibitem{verpillat2011dark}
F.~Verpillat, F.~Joud, P.~Desbiolles, and M.~Gross, \enquote{Dark-field digital
  holographic microscopy for 3D-tracking of gold nanoparticles,} Opt. Express
  \textbf{19}(27), 26,044--26,055 (2011).

\bibitem{gross2003shot}
M.~Gross, P.~Goy, and M.~Al-Koussa, \enquote{Shot-noise detection of
  ultrasound-tagged photons in ultrasound-modulated optical imaging,} Opt.
  Lett. \textbf{28}(24), 2482--2484 (2003).

\bibitem{absil2010photothermal}
E.~Absil, G.~Tessier, M.~Gross, M.~Atlan, N.~Warnasooriya, S.~Suck,
  M.~Coppey-Moisan, D.~Fournier, \enquote{Photothermal
  heterodyne holography of gold nanoparticles,} Opt. Express \textbf{18},
  780--786 (2010).

\bibitem{psota2012comparison}
P.~Psota, V.~L{\'e}dl, R.~Dole{\v{c}}ek, J.~V{\'a}clav{\'\i}k, and
  M.~{\v{S}}ulc, \enquote{Comparison of Digital Holographic Method for Very
  Small Amplitudes Measurement with Single Point Laser Interferometer and Laser
  Doppler Vibrometer,} in \emph{Digital Holography and Three-Dimensional
  Imaging} (Optical Society of America, 2012).

\bibitem{joud2009imaging}
F.~Joud, F.~Lalo{\"e}, M.~Atlan, J.~Hare, and M.~Gross, \enquote{Imaging a
  vibrating object by Sideband Digital Holography,} Opt. Express
  \textbf{17}(4), 2774--2779 (2009).

\bibitem{davenport1958introduction}
W.~Davenport and W.~Root, \emph{An introduction to the theory of random signals
  and noise} (McGraw-Hill New York, 1958).

\bibitem{charriere2006sni}
F.~Charri{\`e}re, T.~Colomb, F.~Montfort, E.~Cuche, P.~Marquet, and
  C.~Depeursinge, \enquote{{Shot-noise influence on the reconstructed phase
  image signal-to-noise ratio in digital holographic microscopy},} Appl.
  Opt. \textbf{45}(29), 7667--7673 (2006).

\bibitem{charriere2007isn}
F.~Charri{\`e}re, B.~Rappaz, J.~K{\"u}hn, T.~Colomb, P.~Marquet, and
  C.~Depeursinge, \enquote{{Influence of shot noise on phase measurement
  accuracy in digital holographic microscopy},} Opt. Express \textbf{15}(14),
  8818--8831 (2007).

\bibitem{verpillat2010digital}
F.~Verpillat, F.~Joud, M.~Atlan, and M.~Gross, \enquote{Digital holography at
  shot noise level,} IEEE J. Disp. Technol. \textbf{6}(10), 455--464
  (2010).

\bibitem{cuche2000spatial}
E.~Cuche, P.~Marquet, and C.~Depeursinge, \enquote{{Spatial filtering for
  zero-order and twin-image elimination in digital off-axis holography},}
  Appl. Opt. \textbf{39}(23), 4070--4075 (2000).

\bibitem{kreis1997suppression}
T.~Kreis and W.~Jueptner, \enquote{{Suppression of the dc term in digital
  holography},} Opt. Eng. \textbf{36}(8), 2357--2360 (1997).

\bibitem{picart2008general}
P.~Picart and J.~Leval, \enquote{General theoretical formulation of image
  formation in digital Fresnel holography,} J. Opt. Soc. Am. A \textbf{25}(7), 1744--1761
  (2008).

\bibitem{verrier2011off}
N.~Verrier and M.~Atlan, \enquote{Off-axis digital hologram reconstruction:
  some practical considerations,} Appl. Opt. \textbf{50}(34), H136--H146
  (2011).

\bibitem{yu2005wavelength}
L.~Yu and M.~Kim, \enquote{Wavelength-scanning digital interference holography
  for tomographic three-dimensional imaging by use of the angular spectrum
  method,} Opt. Lett. \textbf{30}(16), 2092--2094 (2005).

\bibitem{bachor_1998}
H.~A. Bachor, \emph{A guide to experiment in quantum optics} (Wiley, Weinheim,
  New York, 1998).

\bibitem{Newberry1998}
M.~Newberry, \enquote{Measuring the Gain of a CCD Camera,} Axiom Tech Note
  \textbf{1}, 1--8 (1998).



\end{thebibliography}

\end{document}